\documentclass[pra,twocolumn,showpacs,preprintnumbers,amsmath,amssymb]{revtex4}

\usepackage{graphicx}% Include figure files
\usepackage{dcolumn}% Align table columns on decimal point
\usepackage{bm}% bold math

\begin{document}

\title{Two-channel Feshbach physics in a structured continuum}
\author{N. Nygaard}
\author{R. Piil}
\author{K. M\o lmer}
\affiliation{Lundbeck Foundation Theoretical Center for Quantum System Research, Department of Physics and Astronomy, University of Aarhus, DK-8000 {\AA}rhus C, Denmark}

\date{\today}% It is always \today, today,

\begin{abstract}
We analyze the scattering and bound state physics of a pair of atoms in a one-dimensional optical lattice interacting via a narrow 
Feshbach resonance. The lattice provides a structured continuum allowing for the existence of bound dimer states both below and above the continuum bands, with pairs above the continuum stabilized by either repulsive interactions or their center of mass motion. Inside the band the Feshbach coupling to a closed channel bound state leads to a Fano resonance profile for the transmission, which may be mapped out by RF- or photodissociative spectroscopy. We generalize the scattering length concept to the one-dimensional lattice, where a scattering length may be defined at both the lower and the upper continuum thresholds. As a function of the applied magnetic field the scattering length at either band edge exhibits the usual Feshbach divergence when a bound state enters or exits the continuum. Near the scattering length divergences the binding energy and wavefunction of the weakly bound dimer state acquires a universal form reminiscent of those of free-space Feshbach molecules. We give numerical examples  of our analytic results for a specific Feshbach resonance, which has been studied experimentally.
\end{abstract}

\maketitle

\section{Introduction}

An optical lattice is a crystal structure of off-resonant laser light, wherein atoms are trapped due to the optical dipole force. This creates a periodic potential in which the motional states of a single atom exhibits many of the usual textbook solid state characteristics, such as energy band structure, Bloch oscillations and Wannier-Stark ladders~\cite{Dahan1996,Wilkinson1996}. 
In an interacting  many-body system the periodic potential constitutes an ideal setting for realizing a wide range of condensed matter phenomena such as the superfluid to Mott insulator transition~\cite{Greiner2002,Bloch2007}. Likewise controlled atomic collisions in an optical lattice potential is a promising approach to implement quantum gate operations~\cite{Mandel2003}.   

In the context of ultracold atomic collisions a Feshbach resonance arises, when a bound state of a closed scattering channel is coupled to the scattering continuum in the entrance channel~\cite{Kohler2006}.  In an applied magnetic field the separated atoms are described by eigenstates of the hyperfine and Zeeman-Hamiltonian, and the coupling between the different channels is a result of the atomic interaction, which is not diagonal in the hyperfine basis. The position of the resonance is tunable by varying the strength of the applied magnetic field, since the open and closed channels have different magnetic moments, and therefore experience different Zeeman shifts. When the closed channel bound state is tuned to be degenerate with the colliding atoms in the entrance channel the scattering cross section is resonantly enhanced. A Feshbach resonance may thus be used to regulate the atomic interactions, which in the ultracold regime is characterized entirely by the $s$-wave scattering length. In addition, since the Feshbach resonance provides an adiabatic connection between bound dimer states and the continuum, a pair of unbound atoms may be converted into a molecule by slowly ramping the magnetic field across the resonance position. 

To study two-particle physics on a lattice experimentally, a state with either zero or two atoms at each lattice site may be prepared by adiabatic magnetic field sweeps across a Feshbach resonance starting from a pure atomic cloud~\cite{Thalhammer2006,Volz2006}. This prepares a state with at most two atoms in each lattice well, since sites with more than two atoms are eliminated by inelastic collisions. The unbound atoms may be subsequently removed by an optical purification pulse. The sample thus prepared contains only isolated pairs, whose binding can then be controlled by further adjustments of the magnetic field.  

A special situation arises in an optical lattice due to the band structure. The periodic potential leads to a structured continuum with scattering states grouped in energy bands separated by band gaps. The depth of the lattice controls the width of the bands and their mutual separation. In this paper we concentrate on the lowest band, where the band structure then implies that the continuum has both a lower and an upper edge. This has fundamental implications for both the bound states and the scattering states of the system.
A counter-intuitive effect of the lattice band structure is the existence of bound pairs composed of atoms, which repel each other~\cite{Winkler2006,Ospelkaus2006}. Such repulsively bound pairs exist, when the repulsive interaction is strong enough to lift the state of two atoms on the same lattice site out of the continuum and into the band gap, thereby preventing the pair from dissociating and making it a bound state of the system~\cite{Veiga2002,Mahajan2006,Denschlag2006}. In one dimension repulsively bound pairs exist for an arbitrarily small repulsion. 
Near a Feshbach resonance further intriguing two-body effects become possible.   
In particular, the Feshbach resonance gives rise to motionally bound molecules, which are stable only at finite center of mass momentum with respect to the lattice, while decaying if at rest. It is thus possible to tune the location of the scattering resonance at a fixed magnetic field by controlling the molecular motion~\cite{Grupp2007}. 

Here we derive analytical results fully characterizing both the bound and scattering states of an atom pair in an optical lattice near a Feshbach resonance. We represent the lattice physics in a discrete model with nearest neighbor hopping. To describe the Feshbach resonance we adopt a two-channel model with one open channel and one closed channel, which is adequate to describe the physics of both the scattering continuum and the bound states of the system for an isolated resonance~\cite{Kohler2006,Nygaard2006}.
Contrary to~\cite{Grupp2007} we include the background interaction in the entrance channel, which plays an important role both in scattering and in the bound state dynamics. The direct scattering process in the open channel interferes with the resonant scattering proceeding via the closed channel resonance state leading to a Fano profile in the transmission probability. In the bound state spectrum the presence of an additional bound state leads to avoided crossings that are important when considering adiabatic transitions induced by ramping the magnetic field. 

The same fundamental physics of a discrete state embedded in a structured continuum occurs in the Fano-Anderson model, which describes 
autoionizing states in atoms~\cite{Fano1961} and  localized electron states in solids~\cite{Anderson1961}. From a many-body perspective the combination of an optical lattice with resonant control of the atomic interactions makes a variety of spin models attainable~\cite{Carr2005,Zhou2005,Duan2005}.

In a previous paper~\cite{Nygaard2008} we described the basic results of the discrete lattice model of the Feshbach resonance physics, and suggested how the salient features of the system may be probed in experiments. Here we present the full analytic theory of Feshbach scattering in a one-dimensional lattice (Sec.~\ref{sec:Scattering}). We introduce a generalized, one-dimensional scattering length, which is defined at both the upper and the lower  threshold of the lowest continuum band in the lattice. This generalized scattering length shows a characteristic Feshbach divergence when a bound state of the system crosses either band edge, and we identify the magnetic field positions for these divergences as a function of the lattice depth for a specific Feshbach resonance. In Sec.~\ref{sec:bound-states} we consider the bound state solutions of the combined lattice-Feshbach problem and derive analytic expressions for the binding energies and the wavefunctions of the dimer states. We also show that the molecules exhibit cooperative tunneling with a tunneling rate consistent with that measured in experiments~\cite{Foelling2007,Strohmaier2007}. A spectral analysis of the unbound and dimerized atom pair states is presented in Sec.~\ref{sec:spectral-analysis}, where we demonstrate that threshold effects lead to a non-trivial peak structure of the molecular spectral function inside the band. Our discrete lattice model includes only the lowest continuum band and neglects couplings to higher bands in the lattice. In Sec.~\ref{sec:validity} we discuss when the lowest band approximation is valid and how this restriction may be relaxed. We find that a moderate interband coupling only modifies our results on a quantitative level. Throughout we give quantitative results for a specific Feshbach resonance for which our model is valid~\cite{Syassen2007}.

\section{One-dimensional scattering in a lattice}
\label{sec:Scattering}

\subsection{Discrete lattice model}

\subsubsection{Single particle states in the discrete basis}

We consider atoms with mass $m$ in a cubic optical lattice potential
$V_{\rm lat}({\bf x})=\sum_{i=1,2} V^\perp_0 E_{\rm{R}} \sin^2(\pi x_i/a)+V^\parallel_0 E_{\rm{R}} \sin^2(\pi x_3/a)$
created by off-resonant laser light with wavelength $\lambda_{\rm{L}}$. The lattice period is, $a=\lambda_{\rm{L}}/2$, and the
transverse, $V^\perp_0$, and longitudinal depths, $V_0^\parallel$, are measured in units of the photon recoil energy, {\mbox{$E_{\rm{R}}={h^2}/{2m\lambda_{\rm{L}}^2}$}}.  For a cubic lattice the motion separates along the three spatial directions, and the
Hamiltonian, ${\mathcal{H}}_0= -(\hbar^2/2m)\nabla^2+V_{\rm lat}({\bf x})$, is diagonalized by products of three Bloch functions,
$\phi_{nq}(x_i)$, one for each spatial direction $i=1,2,3$, where $q$ is the quasi-momentum in the range $[-\pi/a,\pi/a]$, and $n$ is the band index. 
The energy eigenstates form the usual band structure, splitting the continuum into bands separated by energy gaps that decrease with increasing $n$. The band gaps grow and the widths of the bands decrease, as the lattice is made deeper by increasing the laser intensity. 

We will assume that the lattice depth along the transverse directions, $V^\perp_0$, is much larger than the depth of the longitudinal lattice, $V_0^\parallel$. This allows us to only include the ground state along the two perpendicular
directions, leaving an effectively one-dimensional lattice. As the scattering energy is increased consecutive transverse channels open, however we only consider scattering at energies much smaller than the transverse level splitting, thus justifying the neglect of the transverse degrees of freedom.  A confinement induced resonance occurs at the lower edge of the continuum when the (free space) scattering length $a^{3\rm{D}}$ and the width of the transverse confining potential $a_\perp=\sqrt{2}a(V_0^\perp)^{-1/4}/\pi$ are related by $a_\perp/a^{3\rm{D}}=1.4603\ldots$~\cite{Olshanii1998,Bergeman2003,Peano2005,Kim2005,Yurovsky2005,Yurovsky2006a}. However, this resonance may be avoided by not choosing the transverse confinement too tight, and for the quantitative results we present here, it plays no role.

The Bloch waves are delocalized,
but can be combined into orthonormal localized basis functions, Wannier functions, which in the longitudinal direction take the form
\begin{equation}
  \label{eq:WannierFunction}
  w_{nj}(x_3)=\left(\frac{a}{2\pi}\right)^{1/2} \int_{-\pi/a}^{\pi/a}dq\, e^{iqja}\phi_{nq}(x_3).
\end{equation}
In addition to the band index they are labeled by a site index, $j\in {\mathbb{Z}}$, specifying at which
lattice sites, $x_3=ja$, the Wannier functions are localized. A similar Wannier basis can be constructed in the two transverse directions. When it is not clear from the context, which Wannier basis we refer to, we will add a superscript $\perp$ or
$\parallel$ to the Wannier functions indicating to which lattice direction they
belong.

The dynamics of a single atom in the $n$'th longitudinal band, {\textit{i.e.}} the
kinetic energy and lattice potential, is characterized by
the tunneling amplitude, {\mbox{$J_{nm}\equiv -\langle
w_{nj}|{\mathcal{H}}^{\parallel}_0|w_{nj+m}\rangle$}}, between lattice sites separated by a distance $ma$.
Here ${\mathcal{H}}^{\parallel}_0$ only depends on the
coordinate along the weak lattice direction, and $m$ is an integer. In the transverse directions tunneling can be disregarded, due to the much larger lattice depth. 

In this work we consider only the motion in the lowest band of the lattice, and we disregard tunneling between non-adjacent sites, assuming $|J_{1m}|\ll J_{11}$ for all \mbox{$|m|>1$}. 
In the basis of the Wannier states the single particle Hamiltonian for motion in the longitudinal direction then becomes
\begin{equation}
h_0 = -\sum_j J \left( |w_{1j}\rangle \langle w_{1j+1}|+|w_{1j+1}\rangle \langle w_{1j}|\right),
\label{eq:h0}
\end{equation}
with $J\equiv J_{11}$. We have set the zero of energy to be the expectation value of the energy for an atom in a lowest band Wannier state,
{\textit{i.e.}} $\langle w_{1j}^\parallel|{\mathcal{H}}_0^\parallel|w_{1j}^\parallel\rangle$. 
The Hilbert space is now represented by a discrete basis of the localized Wannier functions, and accordingly we have translated the continuum description of the system into a discrete lattice model. By our definition $E=0$ corresponds to the center of the lowest longitudinal Bloch band, and the single-particle energies of the discrete lattice model are $E(q)=-2J\cos(qa)$.
The approximations made are the omission of higher bands, and the neglect of tunneling beyond nearest neighbor sites. Both of these restrictions may be relaxed in a systematic way~\cite{Gurvitz1993,Piil2007}.

\subsubsection{Two-particle states in the discrete basis}

We use the longitudinal Wannier functions as a basis for the two-particle states on the lattice.
Hence the two-particle wavefunction $\psi(z_1,z_2)$ will denote the
amplitude of finding the atoms in the Wannier orbitals
$w_{mj_1}(x_3)$ and $w_{nj_2}(x_3)$ localized around the discrete positions $z_1=j_1a$ and $z_2=j_2a$. Considering only the lowest Bloch band we have $m=n=1$.

To analyze the motion of the two particles in the lattice we introduce center of mass, $Z=(z_1+z_2)/2$, and relative, $z=z_1-z_2$, coordinates. For the center of mass motion a plane-wave ansatz 
\begin{equation}
\psi(z_1,z_2)=e^{iKZ} \psi_K(z),
\label{sep_ansatz}
\end{equation}
is possible, with the first Brillouin zone for the center of mass momentum, $K$, running from $-\pi/a$ to $\pi/a$. This separation is pertinent, since states with distinct center of mass momenta are not coupled by the interactions or the lattice, and we can thus obtain separate results for the relative motion for each $K$. 
Taking the interaction potential $\hat{U}$ to be purely on-site with strength $U$ (defined in Appendix~\ref{sec:calc-param}), the relative motion of two atoms in the open channel is then described by the Hamiltonian, $H^{\rm{op}}= H_0+U\delta_{z,0}$, where the relative motion lattice Hamiltonian, $H_0=-2J\Delta_z^K+E_K$, contains the discrete Laplacian~\cite{Denschlag2006} 
\begin{equation}
\Delta_z^K \psi_K(z) = -\frac{E_K}{4J} [\psi_K(z+a)+\psi_K(z-a)-2\psi_K(z)],
\label{lattice_laplacian}
\end{equation}
and the center of mass energy, $E_K = -4J \cos\left(Ka/2\right)$.

\subsection{Entrance channel Green's functions}
\label{sec:entr-chann-greens}

The eigenstates of $H_0$ are discrete plane waves, $\langle z|k\rangle=\sqrt{a/2\pi}\exp({ikz})$. The corresponding energies, $\epsilon_K(k)=E_K\cos(ka)$, are the sum of the single-particle energies of atoms with momenta $K/2\pm k$, $\epsilon_K(k)=-2J[\cos(Ka/2+ka)+\cos(Ka/2-ka)]$. The (retarded) Green's function for non-interacting atoms on the lattice is $\hat{G}_K^0(E)=[E-H_0+i\eta]^{-1}$, where $\eta$ is a positive infinitesimal. In momentum space it is diagonal
\begin{equation}
{\mathcal{G}}_K^0(E,k,k') = \langle k|\hat{G}_K^0(E)|k'\rangle = \frac{\delta(k-k')}{E-\epsilon_K(k)},
\label{G0_k}
\end{equation}
while its coordinate space form may be found from
\begin{equation}
G^0_K(E,z) = \int_{-\pi/a}^{\pi/a}  \frac{dk}{2\pi} \frac{ae^{ikz}}{E-E_K\cos(ka)+i\eta}.
\label{G0_int}
\end{equation}
Due to the delta function interactions we only need to consider $G_K^0(E,z)\equiv \langle z|\hat{G}_K^0(E)|0\rangle$.
For energies inside the band ($|E|<|E_K|$) the solution is propagating
\begin{equation}
G^0_K(E,z) = -\frac{i\exp(ip|z|)}{\sqrt{E^2_K-E^2}},
\label{G0_z_in}
\end{equation}
with $pa=\cos^{-1}(E/E_K)$, while the the solution outside the band ($|E|>|E_K|$) falls off exponentially:
\begin{equation}
G^0_K(E,z) = {\rm{sgn}}(E)\frac{\exp(-\kappa|z|)}{\sqrt{E^2-E^2_K}}[-{\rm{sgn}}(E)]^{z/a}.
\label{G0_z_out}
\end{equation}
Here $\kappa a=\cosh^{-1}|E/E_K|$. For energies above the band the sign of $G^0_K(E,z)$ alternates between lattice sites.

The Green's function related to the open channel Hamiltonian, including the interaction, $\hat{U}$, is found by explicitly solving the Dyson equation
\begin{equation}
\hat{G}^U_K(E)=\hat{G}_K^0(E)+\hat{G}_K^0(E)\hat{U}\hat{G}^U_K(E). 
\label{Dyson}
\end{equation}
In coordinate space the result has the simple form
\begin{equation}
G^U_K(E,z) = \frac{G^0_K(E,z)}{1-UG^0_K(E,0)}.
\label{GU_z}
\end{equation}
Defining the diagonal elements of the non-interacting momentum space Green's function to be ${\mathcal{G}}^0_K(E,k)=[E-\epsilon_K(k)]^{-1}$ the interacting open channel Green's function has the momentum space form
\begin{eqnarray}
{\mathcal{G}}_K^U(E,k,k') &=& {\mathcal{G}}_K^0(E,k,k') \nonumber \\
&& + \frac{a}{2\pi} \frac{U{\mathcal{G}}_K^0(E,k){\mathcal{G}}_K^0(E,k')}{1-UG_K^0(E,0)},
\label{GU_k}
\end{eqnarray}
which follows from the matrix elements of (\ref{Dyson}) in the relative momentum eigenbasis, using that for an on-site interaction the matrix element, $\langle k|\hat{U}|k'\rangle=Ua/2\pi$, is constant  for all $(k, k')$.   

\subsection{Single channel scattering}

With the expressions for the Green's functions the formal solution of the Schr{\"o}dinger equation for the relative motion becomes explicit.
Below we derive the transmission probability for open channel collisions and analyze the near threshold scattering in terms of a generalized scattering length. This section introduces the concepts and quantities central to the ensuing discussion of two-channel scattering. We first present results relevant to scattering of distinguishable particles, and then discuss how to treat the case of identical bosons or fermions. 

\subsubsection{Transmission profile}
Choosing the situation with the incident wave entering from the left with quasi-momentum $pa=\cos^{-1}(E/E_K)$, the scattering state of two particles colliding under the influence of the entrance channel potential, $\hat{U}$, is given by the solution of the Lippmann-Schwinger equation
\begin{equation}
\psi_K^{\rm{bg}}(E,z)= e^{ipz}+\sum_{z',z''} \langle z |\hat{G}^0_K(E)|z'\rangle \langle z'|\hat{U}|z''\rangle \psi_K^{\rm{bg}}(E,z'').
\label{psi_E__bg_LS}
\end{equation}
For our on-site interaction of strength, $U$, the scattering wavefunction becomes  
\begin{equation}
\psi_K^{\rm{bg}}(E,z)= e^{ipz}+UG^U_K(E,0)e^{ip|z|},
\label{psi_E_bg}
\end{equation}
where we have used that $G^U_K(E,z)=G^U_K(E,0)e^{ip|z|}$.  
We note that for the contact potential $\psi_K^{\rm{bg}}(E,z)$ equals its asymptotic form 
\begin{equation}
\psi_E^{\rm{bg}}(z)\rightarrow e^{ipz}+f_{\rm{bg}}(E,K)e^{ip|z|} \ \ \ (|z|\rightarrow \infty)
\label{psi_E_bg_asymp}
\end{equation}
everywhere, 
hence we may readily identify the scattering amplitude for the background interaction:
\begin{equation}
f_{\rm{bg}}(E,K)=\frac{UG_K^0(E,0)}{1-UG_K^0(E,0)},
\label{f_bg}
\end{equation}  
which is unit-less in a one-dimensional system.
For the one-dimensional problem and distinguishable particles it is natural to recast this in terms of transmission and reflection amplitudes. These are defined from the left and right asymptotic form of the wavefunction 
\begin{equation}
\psi_K^{\rm{bg}}(E,z)=
\left\lbrace
\begin{array}{cc}
t_{\rm{bg}}(E,K)e^{ipz} & (z\rightarrow \infty), \\
e^{ipz} + r_{\rm{bg}}(E,K)e^{-ipz} & (z\rightarrow -\infty).
\end{array}
\right.
\label{tr_definition}
\end{equation}
By comparing with (\ref{psi_E_bg_asymp}) it follows that the transmission amplitude is $t_{
\rm{bg}}=1+f_{\rm{bg}}$, while the reflection amplitude coincides with the scattering amplitude, $r_{\rm{bg}}=f_{\rm{bg}}$. The resulting transmission probability is
\begin{equation}
T_{\rm{bg}}(E,K) = |t_{\rm{bg}}(E,K)|^2 = \frac{E_K^2-E^2}{E_K^2+U^2-E^2}.
\end{equation}
As one might expect, the transmission vanishes in the limit where $|U|\rightarrow\infty$, and approaches unity as the strength of the on-site interaction is diminished. As usual, the reflection coefficient is $R_{\rm{bg}} = |r_{\rm{bg}}|^2=1-T_{\rm{bg}}$, which is unity at the edges of the band, where propagation ceases. The minimum value of the reflection probability, $U^2/(E_K^2+U^2)$, is attained in the middle of the band, $E=0$.
We emphasize that the transmission is independent of the sign of the interaction. This is a special property of the delta-function potential.  

As usual the scattering amplitude may also be related to the phase shift, $\delta_{\rm{bg}}$, of the scattered wave. In  one dimension  the relation is~\cite{Eberly1965}
\begin{equation}
f_{\rm{bg}}(E,K)=\frac{1}{2}\left(e^{2i\delta_{\rm{bg}}(E,K)}-1\right).
\label{eq:f_and_delta}
\end{equation}
In general there will be two partial waves, corresponding to even and odd solutions of the relative motion problem. For the delta-function interaction the odd partial wave has vanishing phase shift due to symmetry, and the reflection and transmission probabilities are then given by $R_{\rm{bg}}=\sin^2(\delta_{\rm{bg}})$ and $T_{\rm{bg}}=\cos^2(\delta_{\rm{bg}})$, respectively.

\subsubsection{Identical particles}
\label{sec:identical}

The preceding derivation of the one-dimensional scattering in the
lattice assumed that the two colliding atoms are distinguishable. If
the collision involves identical bosons or fermions the scattering
state $\psi^{\rm{bg}}_K$ must be an even or an odd function of the
separation $z$, respectively, to make the wavefunction either
symmetric or anti-symmetric under interchange of the two atoms. For
fermionic atoms in the same internal state this implies that the
wavefunction vanishes for $z=0$ and the contact interaction does not produce any scattering. 

For identical bosons the properly symmetrized scattering state may be written as
\begin{equation}
\psi_K^{\rm{bg}}(E,z)=e^{-ip|z|}+e^{2i\delta_{\rm{bg}}}e^{ip|z|}. 
\label{psi_bg_sym}
\end{equation}
When identical particles collide in one dimension the transmission and
reflection coefficients have no meaning, since the incoming and
outgoing fluxes are identical, however the scattered wave still
experiences a phase shift given by (\ref{eq:f_and_delta}). Even though
the transmission profile calculated above does not play a role in the
scattering of identical bosons, it may nonetheless be mapped out
spectroscopically. Consider transitions from a deeper bound state $i$
in the molecule to an energy eigenstate inside the band induced by
applied RF- or photodissociation fields, which are described by some
transition operator $\hat{T}$. If the initial state is well localized
with respect to the lattice spacing, such dissociative transitions
towards the state (\ref{psi_bg_sym}) occur with a probability
$|\langle \psi_K^{\rm{bg}}(E) |\hat{T}| i \rangle|^2\propto
|\psi_K^{\rm{bg}}(E,z=0)|^2\propto \cos^2 \delta_{\rm{bg}}$. This
expression, in turn, is given by the transmission probability,
$T_{\rm{bg}}(E,K)$, obtained above.

\subsubsection{Scattering length}
When the relative quasi-momentum approaches the center or the edges of the Brillouin zone, the two-particle energies $\epsilon_K(k)$  tend to $\pm |E_K|$ and the scattering amplitude attains a limiting form
\begin{equation}
f_{\rm{bg}}(E,K) \rightarrow -\frac{1}{1- i\kappa a|E_K|/U} \ \ \ {\rm{for}} \ \kappa\rightarrow 0, 
\label{f_bg_lim}
\end{equation}
with $\kappa=k$ at the bottom of the band and $\kappa=\pm\pi/a-k$ at the top of the band. These limits allow a natural definition of a generalized one-dimensional scattering length in the lattice 
\begin{equation}
a_{\rm{bg}}(K)=-\frac{a|E_K|}{U}, 
\label{a_bg}
\end{equation}
from the limit $f_{\rm{bg}}\rightarrow -[1+i\kappa a_{\rm{bg}}]^{-1}$ of the scattering amplitude as $\kappa \rightarrow 0$~\cite{Olshanii1998}. An equivalent definition is through the derivative of the phase shift with respect to the relative momentum 
\begin{equation}
a_{\rm{bg}}=- \lim_{\kappa \rightarrow 0} \frac{\partial \delta_{\rm{bg}}(k,K)}{\partial \kappa},
\end{equation}
in analogy with the usual free space scattering in three dimensions. 
Our lattice scattering length is a generalization of the usual concept, in that it is defined for collision energies at both the lower {\textit{and}} the upper edge of the continuum.
The sign of $a_{\rm{bg}}$ at the upper edge of the continuum is a convention, which will be justified for the two-channel case below. Our generalized scattering length depends on the center of mass motion of the pair, which is a crucial feature in the lattice, especially when we discuss the two-channel case below. 

For $U<0$ a bound state is situated below the continuum, making $a_{\rm{bg}}$ positive as expected. In this case the pole of (\ref{f_bg_lim}) lies along the positive imaginary axis, $k= i|U|/a|E_K|$, and the analytic continuation of the plane wave $\exp(ik|z|)$ is the dying exponential of the bound state.  
Conversely, for a repulsively bound pair state lying above the continuum, $U>0$, and $a_{\rm{bg}}$ is negative. The pole of the scattering amplitude is then at $k=\pm\pi/a+iU/a|E_K|$,  corresponding to a wave function decaying exponentially with the separation between the two atoms and a phase factor $\exp(\pm i\pi |z|/a)$, which alternates between $1$ and $-1$ from one lattice site to the next, in accordance with the behavior of the Green's function for the relative motion at $E>|E_K|$ (\ref{G0_z_out}).     

As $|U|\rightarrow 0$ the scattering length diverges, and the bound state approaches the edge of the continuum. Our intuition from scattering theory in three dimensions about the relation between weakly bound states and the scattering length thus holds both below and above the continuum band.
However, since $U$ is proportional to the free space background scattering length, $a_{\rm{3D}}$, the generalized one-dimensional scattering length and its three-dimensional equivalent are inversely related~\cite{Olshanii1998}.

\subsection{Two-channel scattering}
\label{sec:two-chann-scatt}

\begin{figure}[htbp]
  \begin{center}
    \includegraphics[width=.9\columnwidth]{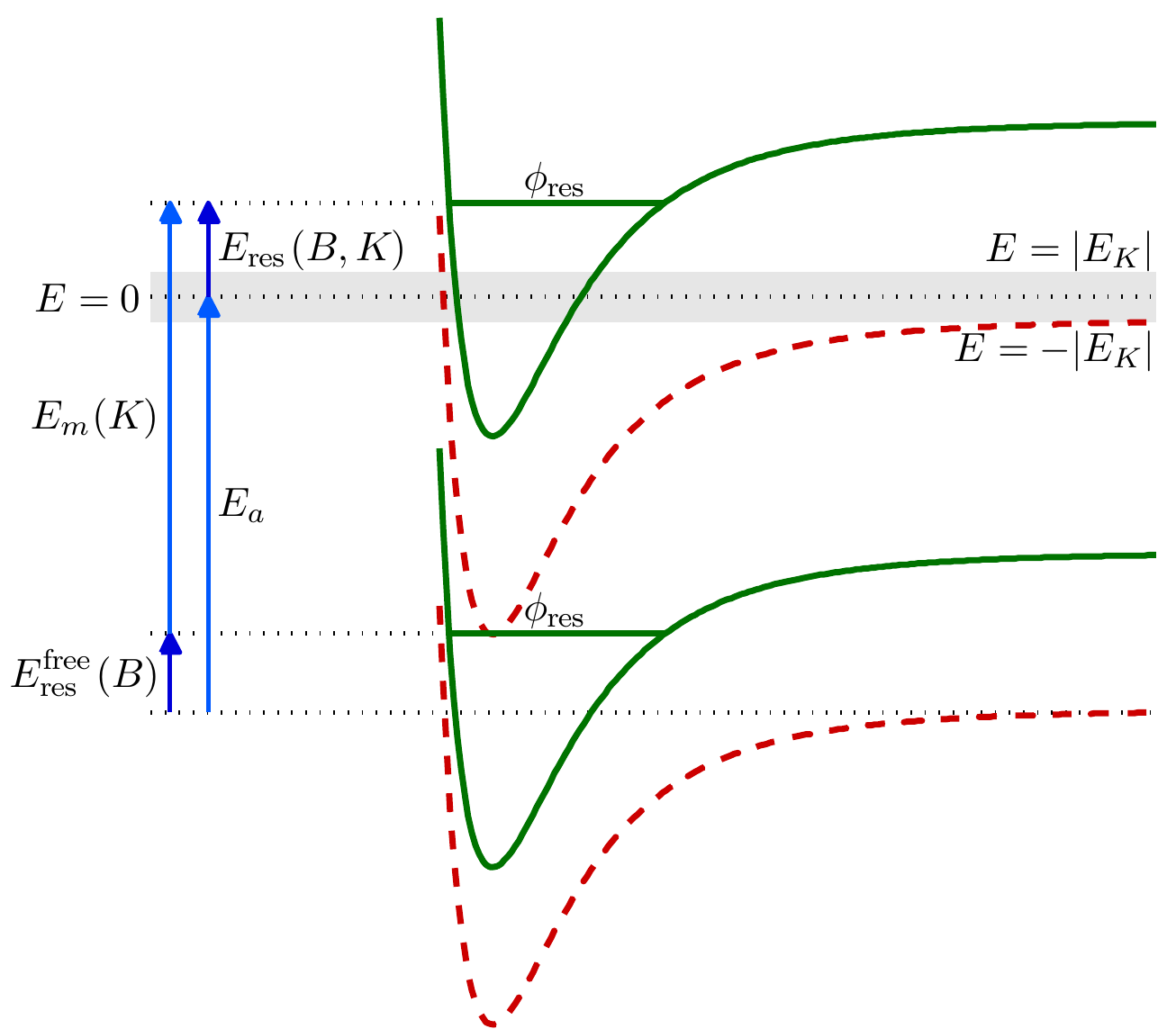} % Add: E=0 $E_{\re th}^{\rm free}$
    \caption{(Color online) Schematic diagram illustrating the difference between two-channel scattering in free space and in an optical lattice. The dashed and solid curves are the open
      and closed channel potentials, respectively. For the lower pair of
      potential curves the lattice is absent, and $E_{\rm res}^{\rm
        free}(B)$ is the energy of $\phi_{\rm res}$ relative to the
      open channel threshold. Applying an optical
      lattice changes the position of the open and closed channel by
      $E_{a}=2(E_1^{\parallel}+2E_1^{\perp})$ and
      $E_m(K)=E_1^{m\parallel}(K)+2E_1^{m\perp}(0)$, respectively,
      where $E_1^{i}$ is the average atomic energy and $E_1^{mi}(K)$ is the
      molecular spectrum of the first Bloch band both along direction
      $i$. Note that the position of the open channel thresholds in the lattice
      depends on the quasi-momentum, $K$, and therefore we choose
      $E=0$ to be the center of the open channel band. Hence, the
      thresholds are located at $\pm |E_K|$. The difference in energy
      between the closed channel bound state and the center of the
      open channel band in the presence of the lattice is
      $E_{\rm{res}}(B,K)=E_{\rm res}^{\rm free}(B)+E_m(K)-E_a=\bar{E}_{\rm{res}}(B)-2J_m\cos (Ka)$. }
  \label{fig:EnergyDiagram}
  \end{center}
\end{figure}

We now consider scattering in the presence of a Feshbach resonance mixing the entrance channel with an energetically closed channel through a coupling, $\hat{W}$. This gives rise to a set of two coupled equations for the relative motion of an atom pair 
 \begin{subequations}
\begin{equation}
  H^{\rm{op}}|\psi^{\rm{op}}_K\rangle + \hat W |\psi_K^{\rm{cl}}\rangle 
  = E |\psi^{\rm{op}}_K\rangle, 
\label{coupled_eq1}
\end{equation}
\begin{equation}
H^{\rm{cl}}|\psi^{\rm{cl}}_K\rangle + \hat W|\psi_K^{\rm{op}}\rangle 
= E |\psi^{\rm{cl}}_K\rangle.
\label{coupled_eq2}
\end{equation}
\end{subequations}
Due to the large splitting between the bound states in the closed
channel the resonance physics is faithfully represented by considering
only a single bound state, the resonance state
$|\phi_{\rm{res}}\rangle$, of $H^{\rm{cl}}$ with an energy
$E_{\rm{res}}(B,K)$ with respect to the center of the open channel
band. This amounts to a single-pole approximation for the closed
channel Green's function
\begin{equation}
\hat{G}_K^{\rm{cl}}(E,B) \approx  \frac{|\phi_{\rm{res}}\rangle \langle \phi_{\rm{res}}|}{E-E_{\rm{res}}(B,K)}.
\label{G_cl_approx}
\end{equation}
The resonance state energy $E_{\rm{res}}(B,K)$ is composed, in part, of the free
space resonance energy $E_{\rm{res}}^{\rm free}(B)$, which is linearly tunable with an applied
magnetic field, due to the difference $\Delta\mu$ between the
separated atoms and the closed channel magnetic moments, and, in part, of the
displacement of the closed channel relative to the open channel Bloch
band as illustrated in Fig. \ref{fig:EnergyDiagram}. At all
magnetic field strengths we take the zero of energy to be the center
of the open channel continuum band. In free space the bare resonance energy
can be related to the magnetic field position of the zero energy
resonance, $B_0^{\rm free}$, as outlined in Appendix~\ref{Resonance_shift}.

As a function of continuous position variables the longitudinal lattice potential for each atom is $V^{\parallel}_0E_{\rm{R}}\sin^2(\pi x_{3i}/a)$ with $i=1,2$, and
therefore the combined lattice potential for a pair of atoms in the
resonance state is
\begin{eqnarray}
V^{\parallel}_{\rm{lat}}(x_{31},x_{32})&=& 2V^{\parallel}_0E_{\rm R}\sin^2\left(\frac{\pi X_3}{a}\right)\cos^2\left(\frac{\pi\delta x_{3}}{2a} \right)
\nonumber \\
&& +2V^{\parallel}_0E_{\rm R}\cos^2\left(\frac{\pi X_3}{a}\right)\sin^2\left(\frac{\pi\delta x_{3}}{2a} \right),
\label{V_lat_mol}
\end{eqnarray}
where we have introduced the center of mass, $X_3=(x_{31}+x_{32})/2$, and relative, $\delta x_3=x_{31}-x_{32}$, coordinates.  
Since the size of the closed channel bound state is much smaller than
the lattice spacing, we may consider the limit of
$\delta x_{3} \ll a$ for this state. Consequently, the resonant state has
a center of mass motion corresponding to a lattice potential twice as
deep as that governing the motion of the individual atoms, while 
its mass is the sum of the two atomic masses. This leads
to a tunneling rate for the resonant state, $J_m$, which is much reduced in
comparison with the rate at which the free atoms tunnel through the
lattice, $J_m \ll J$. This bare molecular tunneling leads to a modulation of the resonance energy with the center of mass momentum of the pair, which we may parametrize as
\begin{equation}
E_{\rm{res}}(B,K)=\bar{E}_{\rm{res}}(B)-2J_m\cos (Ka).
\label{eq:E_res_K}
\end{equation}
The band averaged resonance energy, $\bar{E}_{\rm{res}}(B)$, incorporates the relative energy shifts of the atomic and molecular bands in the lattice.

\subsubsection{Transmission profile}

With the explicit expression for the closed channel Green's function
(\ref{G_cl_approx}) the closed channel part of the scattering problem
(\ref{coupled_eq2}) may be solved formally:
\begin{equation}
  \label{eq:psi_cl_res}
  |\psi_K^{\rm{cl}}(E)\rangle = \frac{|\phi_{\rm{res}}\rangle {\mathcal{W}}\psi^{\rm{op}}_K(0) }{E-E_{\rm{res}}(B,K)},
\end{equation} 
where we have introduced a coupling matrix element, $\langle z, \,
{\rm{op}} |\hat{W}|\phi_{\rm{res}}\rangle =
{\mathcal{W}}\delta_{z,0}$, incorporating the structure of the
resonance state. For a typical resonance in alkali collisions the
coupling arises due to the difference between the triplet and singlet
molecular potentials, and it is therefore inherently of short range
compared with the lattice spacing. It is therefore reasonable that the coupling only acts between atoms localized at the same lattice site. 

The open channel component of the scattering state in the presence of
the resonance may then be expressed as
\begin{equation}
\psi_K^{\rm{op}}(E,z) = t_{\rm{bg}}e^{ipz}\left[ 1+\frac{\Sigma_M(E,K)}{E-E_{\rm{res}}(B,K)-\Sigma_M(E,K)}
\right]
\label{psi_op_res}
\end{equation}
for all $z\geq 0$. Again the relative quasi-momentum of the scattering atoms is $pa=\cos^{-1}(E/E_K)$, and $\Sigma_M(E,K) = \langle \phi_{\rm{res}}|\hat{W}\hat{G}^U_K(E)\hat{W}|\phi_{\rm{res}}\rangle$ is the molecular self-energy, which for the onsite coupling may be written as: $\Sigma_M(E,K)={\mathcal{W}}^2G^U_K(E,0)$.

\begin{figure}[htbp]
\begin{center}
   \includegraphics[width=\columnwidth]{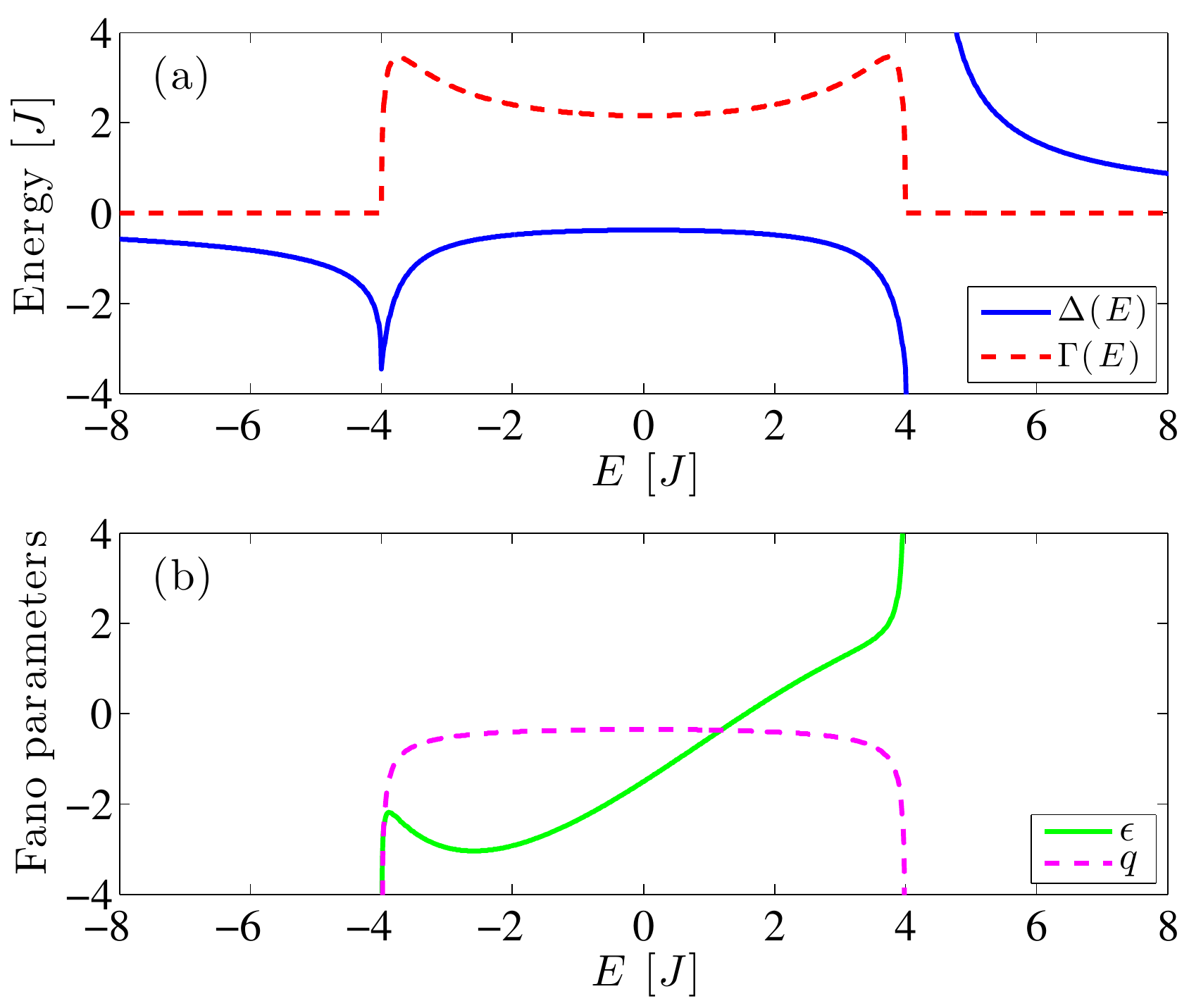} 
\caption{(Color online) The shift, $\Delta(E,K)$ and width, $\Gamma(E,K)$ of the resonance for $K=0$ where the band edges are at $E=\pm 4J$ (a). The real part of the molecular self-energy is continuous across the band edges and has a pole at the uncoupled bound state energy, $E_b^0$ (see section~\ref{sec:bound-states}). The Fano parameters, $\epsilon$ and $q$, parametrize the shape of the transmission resonance (b). Here $\bar{E}_{\rm{res}}=2J$.}
\label{fig:shift_and_width}
\end{center}
\end{figure}

From the form of the scattered wave (\ref{psi_op_res}) the total transmission coefficient including both background and resonant contributions  is found to be of the Fano form~\cite{Fano1961}
\begin{equation}
T(E,K)=T_{\rm{bg}}(E,K)\frac{(\epsilon+q)^2}{\epsilon^2+1}.
\label{Fano}
\end{equation}
The Fano parameters $\epsilon=2(E-E_{\rm{res}}-\Delta)/(\hbar\Gamma)$ and $q=2\Delta/\hbar\Gamma$ depend on the resonance shift and width functions, which are related to the real and imaginary parts of the molecular self-energy, 
\begin{equation}
\Sigma_M(E,K) = \Delta(E,K)-i\hbar\Gamma(E,K)/2, 
\label{sigma_cl}
\end{equation}
where the real part 
\begin{equation}
\Delta(E,K) = -\frac{U{\mathcal{W}}^2}{E_K^2+U^2-E^2} \ \ \ (|E|<|E_K|),  
\label{eq:delta}
\end{equation}
describes the self consistently determined shift of the resonance position from $E_{\rm{res}}$ to $E^\star_{\rm{res}}=E_{\rm{res}}+\Delta(E^\star_{\rm{res}},K)$, and
\begin{equation}
\label{eq:gamma}
\hbar\Gamma(E,K) = \frac{2{\mathcal{W}}^2\sqrt{E_K^2-E^2}}{E_K^2+U^2-E^2}  \ \ \ (|E|<|E_K|). 
\end{equation}
is the decay width of the resonance state due to the coupling to the continuum.
Outside the continuum $\Gamma(E,K)=0$, and the molecules are stable against dissociation. The real part of the molecular self-energy outside the band
\begin{equation}
\Delta(E,K) = \frac{{\rm{sgn}}(E){\mathcal{W}}^2}{\sqrt{E^2-E_K^2}-U{\rm{sgn}}(E)}  \ \ \ (|E|>|E_K|),  
\end{equation}
determines the energies of the bound states for a given resonance energy as described in Section~\ref{sec:bound-states}. The width and shift of the resonance are illustrated in Fig.~\ref{fig:shift_and_width} along with the Fano parameters for $K=0$. Throughout the paper we illustrate our analytical results for a set of parameters where $U=1.4J$ and ${\mathcal{W}}=2.2J$. As outlined in Appendix~\ref{sec:calc-param} this corresponds to $^{87}$Rb atoms  in a lattice with longitudinal and transverse depths of $E_{\rm{R}}$ and 30$E_{\rm{R}}$, respectively, in an applied magnetic field tuned close to  the Feshbach resonance near 414 G~\cite{Syassen2007}. 

We note that $q=-U/\sqrt{E_K^2-E^2}$ is independent of the coupling, ${\mathcal{W}}$. It is related to the background transmission probability through the relation $T_{\rm{bg}}=1/(1+q^2)$, which is a consequence of inversion symmetry in the problem~\cite{Nockel1994}. 

\begin{figure*}[htbp]
\begin{center}
   \includegraphics[width=\textwidth]{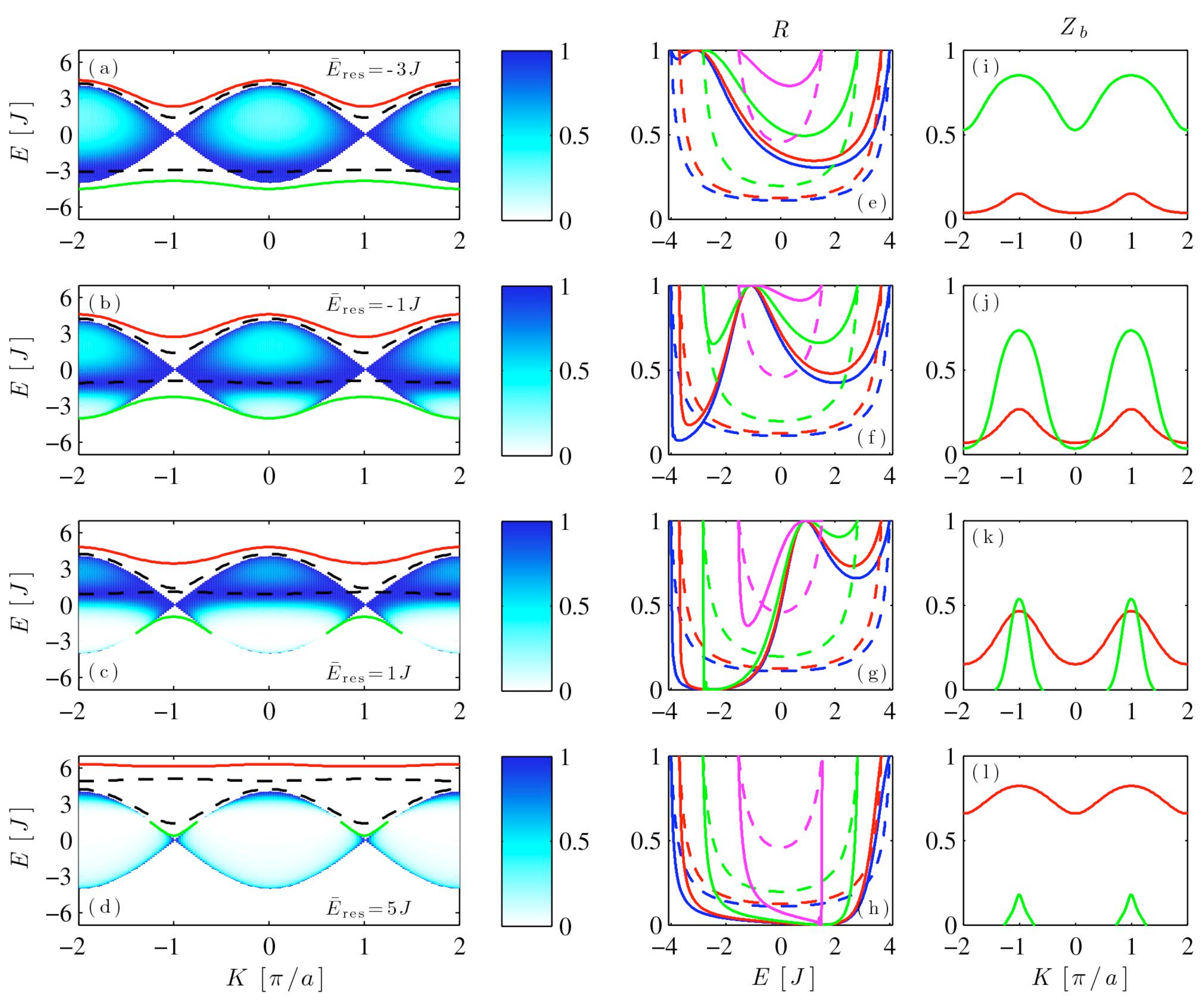} 
\caption{(Color online) (a)-(d) Scattering properties and bound states in an extended zone scheme at different values of the resonance energy. The colormap inside the band illustrates the reflection probability, $R(E,K)$, ranging from 0 (white) to 1 (dark blue). The solid lines are the bound state energies of the coupled system, while the dashed curves indicate the diabatic, uncoupled bound states with the nearly horizontal line giving $E_{\rm{res}}$ and the other corresponding to $E_b^0$ (\ref{eq:Eb0}). 
(e)-(h) Reflection coefficient (solid lines) and the background reflection probability, $R_{\rm{bg}}$ (dashed lines) for $K=$0, 0.25, 0.5, 0.75 $\pi/a$ (from the bottom to the top at $E=0$). (i)-(l) Closed channel population of the bound states, Eq. (\ref{Z}). Plots are shown for resonance energies $\bar{E}_{\rm{res}}=-3J$ (a,e,i), -$J$ (b,f,j), $J$ (c,g,k), and $5J$ (d,h,l).}
\label{fig:R_vs_K}
\end{center}
\end{figure*}

\begin{figure}[htbp]
\begin{center}
   \includegraphics[width=\columnwidth]{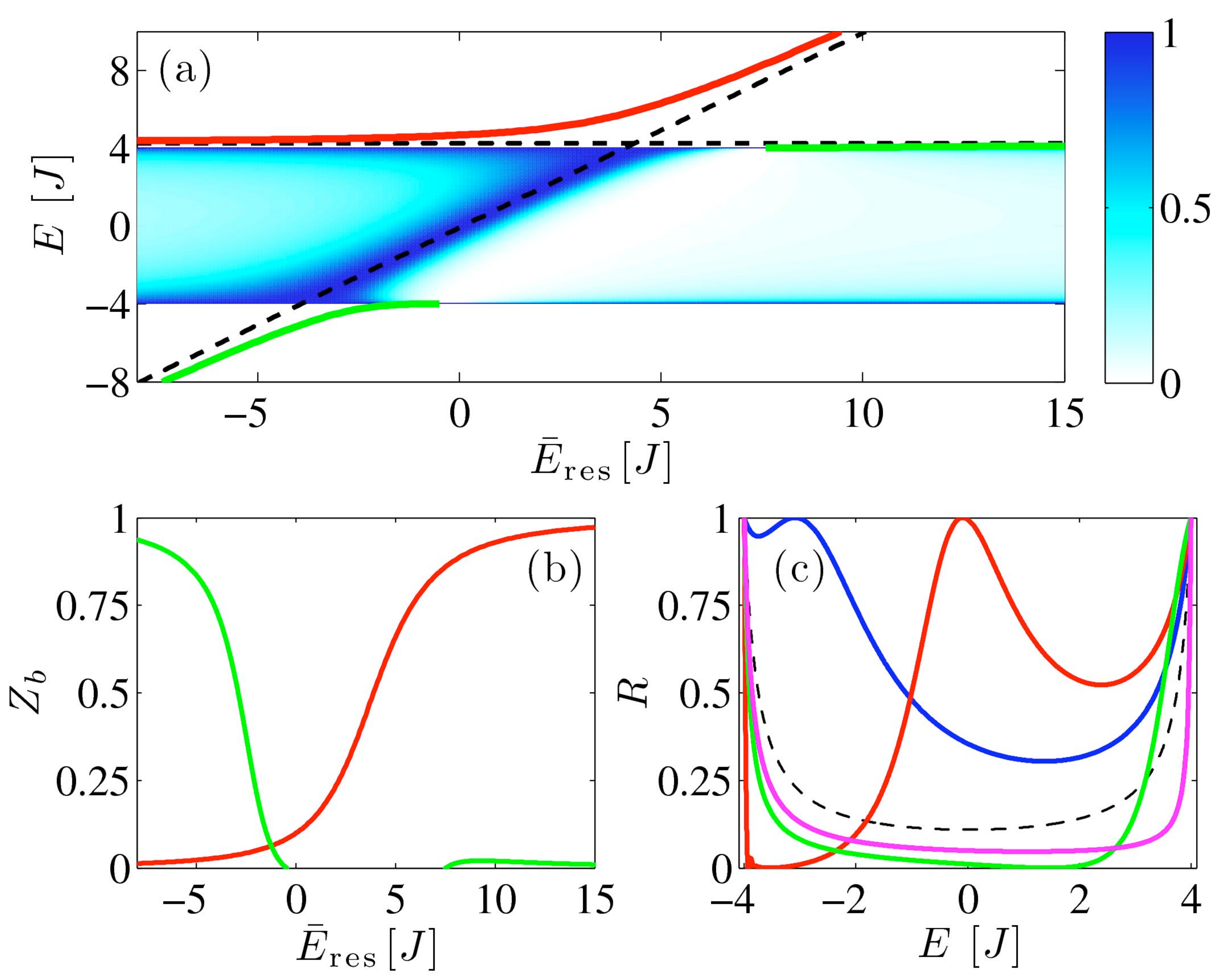} 
\caption{(Color online) (a) Bound state energies (solid lines) and reflection coefficient across the continuum band (colormap) as a function of $\bar{E}_{\rm{res}}$ for a fixed $K=0$. The horizontal dashed line is the repulsively bound pair state existing above the band for ${\mathcal{W}}=0$, while the diagonal dashed curve is the band averaged resonance energy, $\bar{E}_{\rm{res}}$. (b) Closed channel population of the bound states. (c) Cuts showing the reflection probability at $\bar{E}_{\rm{res}}=$ 0, -3, 10, and $5J$ (solid lines from top to bottom at $E=0$). The dashed curve indicate the reflection profile in the absence of the coupling.}
\label{fig:R_vs_Eres}
\end{center}
\end{figure}

The reflection maximum of the Fano profile occurs at
$E=E_{\rm{res}}(B,K)$ and is nearly independent of the center of mass motion,
due to the weak tunneling of the closed channel molecules. This point of
total reflection only occurs, when the bare resonance state lies
inside the continuum. Just as in the non-resonant case the reflection
probability is unity at the band edges. Hence the reflection profile
has two minima, when $|E_{\rm{res}}|<|E_K|$. When the bare resonance
state lies outside the band, $R(E,K)$ has a single minimum. The
two-channel reflection profiles are plotted in
Figs. \ref{fig:R_vs_K}(a)-(d) and \ref{fig:R_vs_Eres}(a) as colormaps
and in Figs. \ref{fig:R_vs_K}(e)-(h) and \ref{fig:R_vs_Eres}(c) as
solid lines. For the parameters used in the quantitative calculations throughout this paper $J_m/J \approx 0.2$, and hence the resonance energy appears nearly independent of $K$ in Figs. \ref{fig:R_vs_K}(a)-(d).

The condition for total transmission is $\epsilon q=1$, corresponding to $E=E_{\rm{res}}(B,K)-{\mathcal{W}}^2/U$. As $E_{\rm{res}}$ is ramped upwards (downwards) starting below (above) the band, this point of destructive interference between the direct and resonant scattering amplitudes first appears at the magnetic field, where a bound state of the coupled channels system enters the continuum to become a scattering resonance. As we shall see below, this occurs at two critical values where $E_{\rm{res}}=\pm|E_K|+{\mathcal{W}}^2/U$ - exactly the points delineating the region, where the condition $\epsilon q=1$ can be satisfied. The reflection profile will thus only have both a Fano maximum and a Fano minimum if the conditions $|E_{\rm{res}}|<|E_K|$ and $|E_{\rm{res}}-{\mathcal{W}}^2/U|<|E_K|$ are simultaneously met.     

Neglecting the background interaction corresponds to the limit $T_{\rm{bg}}\rightarrow 1$, or equivalently $q\rightarrow 0$. In this case the resonance shift vanishes and the reflection probability assumes a simple Breit-Wigner-like profile~\cite{Grupp2007}
\begin{equation}
  \label{eq:Rlimit}
R(E,K) \stackrel{U\rightarrow 0} {\longrightarrow} \frac{\hbar^2\Gamma_0^2(E,K)/4}{[E-E_{\rm{res}}(B,K)]^2+\frac{\hbar^2\Gamma_0^2(E,K)}{4}},
\end{equation}
where $\hbar\Gamma_0(E,K)=2{\mathcal{W}}^2/\sqrt{E_K^2-E^2}$.
The opposite limit of $|U|\rightarrow \infty$ on the other hand corresponds to $T_{\rm{bg}} \rightarrow 0$ and $q\rightarrow \infty$, giving rise to a Breit-Wigner-like form of the transmission coefficient
\begin{equation}
  \label{eq:Tlimit}
T(E,K) \stackrel{|U|\rightarrow \infty} {\longrightarrow} \frac{\hbar^2\Gamma_\infty^2(E,K)/4}{[E-E_{\rm{res}}(B,K)-\Delta_\infty]^2+\frac{\hbar^2\Gamma_\infty^2(E,K)}{4}},
\end{equation}
corresponding to a dip in the reflection probability at the resonance, which now occurs at $E=E_{\rm{res}}(B,K)+\Delta_\infty$. The resonance shift, $\Delta_\infty=-{\mathcal{W}}^2/U$, is independent of energy and momentum, while the width is given by $\hbar\Gamma_\infty(E,K)=2{\mathcal{W}}^2\sqrt{E_K^2-E^2}/U^2$. Both the shift and the width approach zero as the strength of the onsite interaction is increased.

\subsubsection{Spectroscopic probing of the Fano profile}

The observations regarding indistinguishable particles in section~\ref{sec:identical} can be directly carried over to the situation with more than one scattering channel. Again, the transmission coefficient $T(E,K)$ has no straightforward meaning in the scattering of identical bosons. But following Fano~\cite{Fano1961} it is clear that the transition probability from an initial state $i$ to the stationary scattering state $\psi_K^{\rm{op}}$ under the action of some transition operator $\hat{T}$ is $|\langle \psi_K^{\rm{op}}|\hat{T}|i\rangle|^2\propto T(E,K)$. Therefore, the resonance profile is accessible spectroscopically by scanning the energy of the final state across the continuum. Starting from a deeper bound state of the system the dissociated atom signal will have the same energy variation as $T(E,K)$.

\subsubsection{Scattering length}

The scattering amplitude in the coupled channels case consists of a background and a resonant contribution, 
$f(E,K)=f_{\rm{bg}}(E,K)+f_{\rm{res}}(E,K)$, where the resonant part
\begin{equation}
f_{\rm{res}}(E,K)=\frac{{\mathcal{W}}^2G_K^0(E,0)/[1-UG_K^0(E,0)]^2}{E-E_{\rm{res}}(B,K)-\Sigma_M(E,K)}
\label{f_res}
\end{equation}
is obtained by comparing (\ref{psi_op_res}) with $e^{ipz}(1+f_{\rm{bg}}+f_{\rm{res}})$ for $z\rightarrow \infty$. The phase shift is related to the scattering amplitude in the same way as in the single-channel case.

As in the case of single-channel scattering we define the one-dimensional scattering length, $a^\pm_{1{\rm{D}}}(K)$, from the limit of the scattering amplitude, $f\rightarrow -[1+i\kappa a_{1{\rm{D}}}^{\pm}]^{-1}$, at the extrema of the relative energy, $\kappa\rightarrow 0$. After some algebra we arrive at the simple expression
\begin{equation}
a_{1{\rm{D}}}^{\pm}(B,K) = a_{\rm{bg}}(K) \left[ 1+\frac{{\mathcal{W}}^2/U}{E_{\rm{res}}(B,K)-{\mathcal{W}}^2/U\mp |E_K|} \right] 
\label{a_1D}
\end{equation} 
for the scattering length, with the upper and lower sign corresponding to the top  and the bottom edge of the continuum, respectively.
This lattice scattering length is of the standard Feshbach form
\cite{Moerdijk95}, and reduces to the non-resonant expression (\ref{a_bg}), when the coupling ${\mathcal{W}}$ is set to zero. 
However, the generalized, magnetic field dependent scattering length at the top and the bottom of the continuum differ, in that they do not diverge at the same value of the resonance energy. The motivation for the chosen sign convention of the scattering length at the upper band edge is guided by the rationale that the 
two scattering lengths then become equal in the limit of a deep lattice, where $|E_K|$ goes to zero.  

\begin{figure}[htbp]
\begin{center}
   \includegraphics[width=\columnwidth]{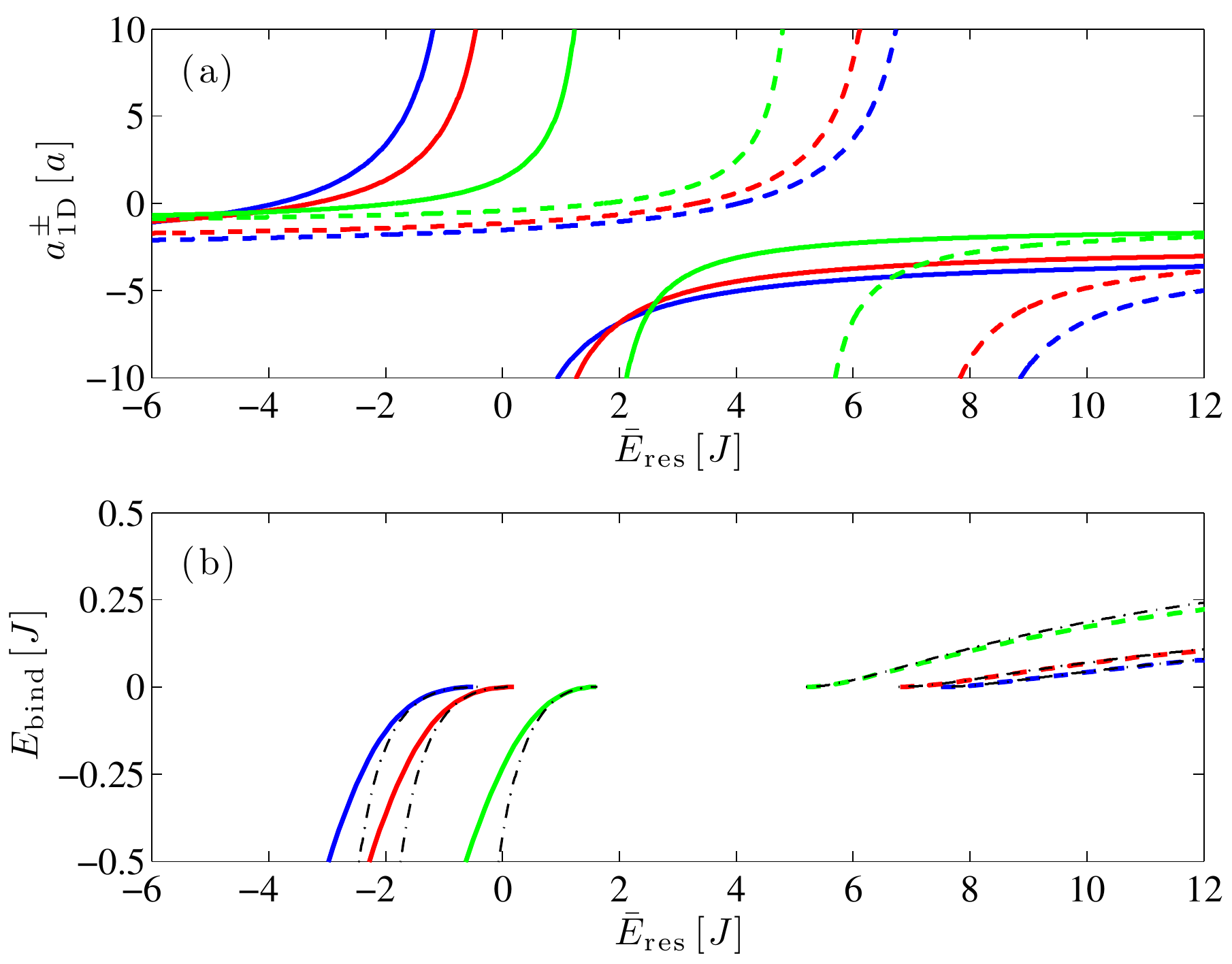} 
\caption{(Color online) (a) Generalized one-dimensional scattering length in the lattice, Eq.~(\ref{a_1D}). Solid lines show $a_{1{\rm D}}^-$ at $K=$ 0, 0.4, and 0.7 $\pi/a$ (from the left to the right), while the dashed lines correspond to $a_{1{\rm D}}^+$ at the same $K$-values (increasing from the right to the left). (b) The corresponding binding energies below the band (solid lines) and above the band (dashed lines) and the limiting form (\ref{E_bind_universal}), valid near the Feshbach resonances in $a_{1{\rm D}}^{\pm}$ (thin dot-dashed lines).}
\label{fig:a_1D}
\end{center}
\end{figure}

\begin{figure}[htbp]
  \begin{center}
    \includegraphics[width=\columnwidth]{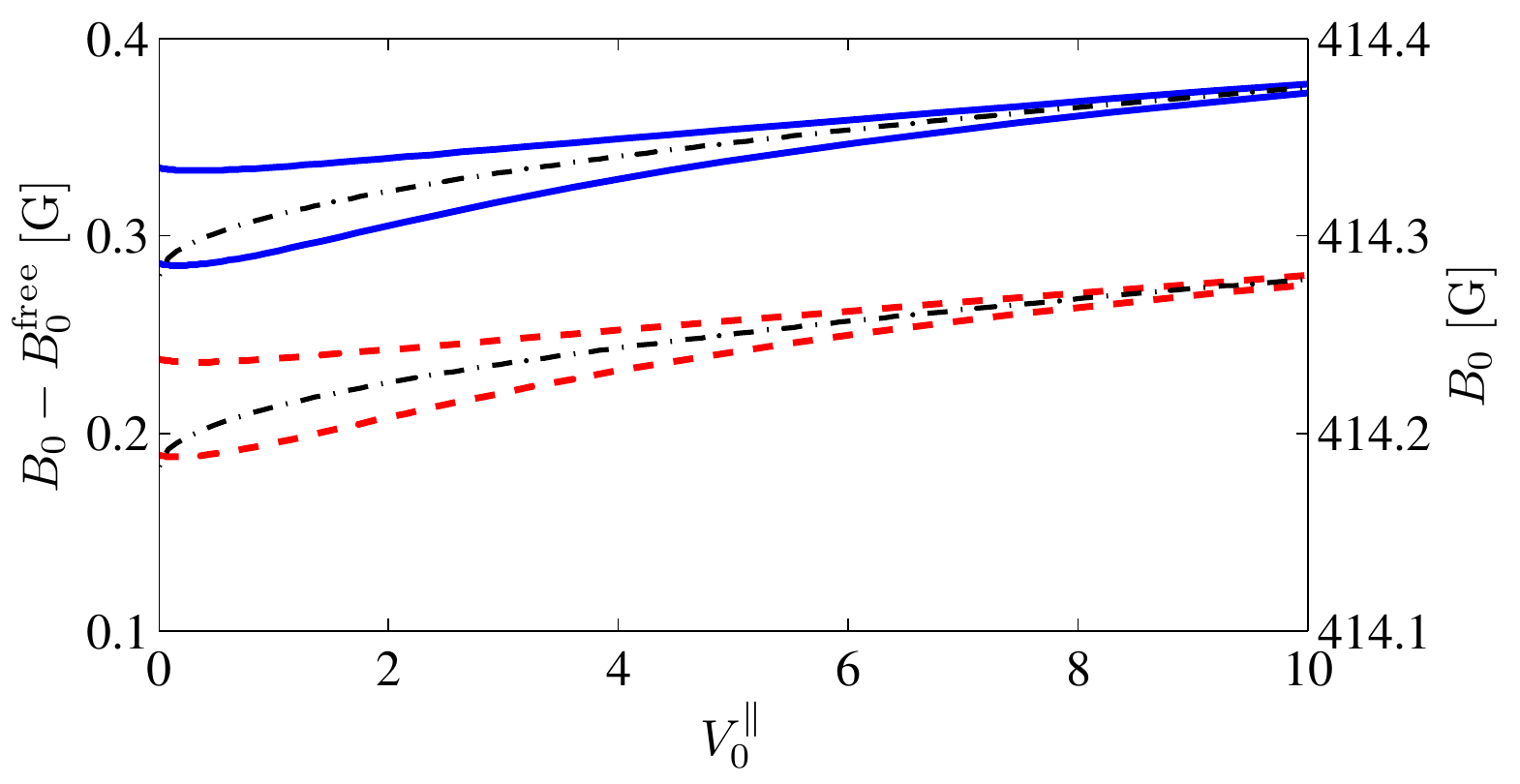} 
    \caption{(Color online) Magnetic field position of the scattering length
      divergences as a function of the lattice strength for
      $K=0$. $B_0^+$ and $B_0^-$ are plotted for $V_0^{\perp}=15$
      (dashed curves) and $V_0^{\perp}=30$ (solid curves). In the example $\Delta \mu$ is positive so both
      $B_0^+$-curves lie above their respective $B_0^-$-curves. The
      left axis shows $B_0^\pm$ relative to the Feshbach
      resonance position in the absence of the lattice, and the right axis shows the actual magnetic
      field position for the specific Feshbach resonance considered~\cite{Syassen2007}. The black dash-dotted curves are given by
      Eq. (\ref{eq:AsymptoticB}).}
  \label{fig:B0}
  \end{center}
\end{figure}

It is convenient to define a pair of detunings
$\delta^{\pm}(B,K)\equiv E_{\rm{res}}(B,K)-{\mathcal{W}}^2/U \mp
|E_K|$, such that the one-dimensional scattering length
diverges at the magnetic field strengths, $B_0^{\pm}$, where
$\delta^\pm(B_0^{\pm},K)=0$.  We thus get two resonance positions, one
for scattering near the upper band edge, and one for scattering near the lower band edge, as shown in 
Fig. \ref{fig:a_1D}. We caution that the relative position of
$B_0^+$ and $B_0^-$ depends on the sign of the magnetic moment
difference, $\Delta\mu$. If $\Delta\mu>0$ ($\Delta\mu<0$), then
$B_0^+$ lies above (below) $B_0^-$. The lattice resonance positions $B_0^+$ and $B_0^-$ are shown in
Fig. \ref{fig:B0} as function of the lattice depth,
$V_0^{\parallel}$. They are separated by
$2|E_K|/\Delta\mu$ and approach each other as $V_0^{\parallel}$ is increased, since $|E_{K}|$
decreases when the lattice becomes deeper. Furthermore, in the limit where both the transverse and the longitudinal lattice depths are
large, \emph{i.e.}, $V_0^{\perp},V_0^{\parallel}\gg 1$, the Wannier orbitals approach harmonic oscillator states, and we have that 
${\mathcal{W}}^2/U\to \Delta\mu\Delta B$, while $E_a\to
2\sqrt{V_0^{\parallel}}E_{\rm R}+4\sqrt{V_0^{\perp}}E_{\rm R}$ and
$E_m(K)\to \sqrt{V_0^{\parallel}}E_{\rm R}+2\sqrt{V_0^{\perp}}E_{\rm
  R}$. This gives the following asymptotic behavior of the magnetic
field resonance position
\begin{equation}
  \label{eq:AsymptoticB}
  B_0^{\pm} \rightarrow B_{0}^{\rm{free}} + \Delta B[1-f(y)]  + \frac{E_{\rm
      R}}{\Delta\mu}\sqrt{V_0^{\parallel}}  + \frac{2E_{\rm{R}}}{\Delta\mu}\sqrt{V_0^\perp}
\end{equation}
where $B_{0}^{\rm{free}}$ is the Feshbach resonance position in the absence of the lattice, and the dimensionless function $f(y)$ is defined in Appendix \ref{Resonance_shift}. The asymptotic
behavior is indicated by the black dash-dotted curves in
Fig. \ref{fig:B0}.

As shown in Sec.~\ref{sec:binding_energy} the divergence of the lattice scattering length is associated with a bound state entering or leaving the continuum. Hence the lattice gives rise to a $K$-dependent shift of the magnetic field threshold for the formation of two-body bound states away from the resonance position in free space. This fact, illustrated in Fig. \ref{fig:a_1D}, was also observed by Orso {\textit{et al.}}~\cite{Orso2005} who solved the two-body physics in a one-dimensional optical lattice within a single-channel model. 
The zero-crossing of the scattering length occurs for $E_{\rm{res}}=\pm |E_K|$, which corresponds to magnetic field values, $B_0^{\pm}-{\mathcal{W}}^2/U\Delta\mu$, shifted from the resonance positions.

\section{Bound states}
\label{sec:bound-states}

The bound state solutions of the coupled-channels problem have
energies, $E_b$, coinciding with the poles of the scattering amplitude
$f(E,K)$. The pole locations satisfy the equation
\begin{equation}
\left[\sqrt{E_b^2-E_K^2}-U{\rm{sgn}}(E_b)\right]=\frac{{\mathcal{W}}^2{\rm{sgn}}(E_b)}{E_b-E_{\rm{res}}(B,K)},
\label{E_bind}
\end{equation}
and can be determined as a subset of the roots of a quartic
polynomial. The bound state energies are plotted as the solid lines in Figs.~\ref{fig:R_vs_K}(a)-(d) and \ref{fig:R_vs_Eres}(a). If $\mathcal{W}=0$ the entrance channel bound state energy is 
\begin{equation}
E^0_b={\rm sign}(U)\sqrt{U^2+E_K^2}
\label{eq:Eb0}
\end{equation}
with a wavefunction proportional to
$G^0_K(E_b,z)$. This single-channel bound state is indicated by the sinusoidal dashed line in Fig~\ref{fig:R_vs_K}(a)-(d) and the horizontal dashed line in Fig.~\ref{fig:R_vs_Eres}(a). 

An important consequence of the structured continuum in the lattice potential is that the stability of a molecular bound state may depend on its center of mass momentum. For a given magnetic field, the scattering resonance only exists for a range of center of mass motion Bloch states, as shown in Fig.~\ref{fig:R_vs_K}(c)-(d)~\cite{Grupp2007}. Outside this range there is instead a true bound state of the system. Such motionally bound states are stabilized by their kinetic energy, which displaces the pair state into the band gap for the relative motion, where the pair cannot disintegrate. Hence, for a fixed magnetic field, where a molecule at rest is unstable, accelerating the lattice provides a new avenue for tuning the Feshbach resonance.

\subsection{Molecular tunneling}

The bound state pairs can tunnel together through the lattice with an effective tunneling rate, which depends on their binding energy. We now determine the tunneling matrix element of the pairs directly from their dispersion relation.
   
In the absence of atomic tunneling processes, $E_K=0$, and the energies of the bound states are given by
\begin{equation}
E_b^{J=0} = \frac{\bar{E}_{\rm{res}}(B)+U}{2}\pm \sqrt{\frac{[\bar{E}_{\rm{res}}(B)-U]^2}{4}+{\mathcal{W}}^2}.
\label{Ebound_no_tunneling}
\end{equation}
With the inter-channel coupling turned off these zero tunneling bound state energies reduce to the uncoupled pair state energies for $J=0$; $\bar{E}_{\rm{res}}$ and $U$. With a small non-vanishing single-particle tunneling amplitude we may expand (\ref{E_bind}) to lowest order in $|E_K/E_b|$ and find 
\begin{equation}
E_b \approx E_b^{J=0}+\frac{E_K^2}{2E^{J=0}_b} \ \ \ (|J|\ll |E_b|).
\end{equation} 
Comparing the momentum dependent  part of the bound state energy with the dispersion relation for non-interacting pairs, $E_{\rm{pair}}=-2J_{\rm{pair}}\cos(Ka)$, we see that in the limit of a deep lattice the bound atom pairs tunnel through the lattice with an effective tunneling rate 
\begin{equation}
J_{\rm{pair}} = -\frac{2J^2}{E_b^{J=0}},
\end{equation}
as one would expect from arguments based on second order perturbation theory~\cite{Micnas1990,Kuklov2003,Duan2003,Petrosyan2007,Miyakawa2007}. We emphasize that the molecule dispersion depends on the applied magnetic field. Cooperative tunneling of atom pairs consistent with an effective pair hopping matrix element given by $J_{\rm{pair}}$ has been demonstrated experimentally by measurement of the frequency of population transfer in a double well~\cite{Foelling2007} and by demonstrating that the transport properties of an attractive Fermi gas in an optical lattice depends strongly on the formation of local pairs~\cite{Strohmaier2007}.

\subsection{Binding energy}
\label{sec:binding_energy}

Instead of the bound state energy it is often instructive to consider the {\textit{binding energy}}, defined as the distance of the bound state energy from the (upper or lower) band edge:
\begin{equation}
E_{\rm{bind}}  = E_b -{\rm{sgn}}(E_b)|E_K|.
\label{E_binding}
\end{equation}
The binding energy is positive for a bound state lying above the band and negative for a bound pair below the continuum. From Fig.~\ref{fig:R_vs_K}(a)-(d) we observe that the binding energy increases with increasing $|K|$ in agreement with the single-channel calculation in~\cite{Orso2005}, which only discussed bound states with negative binding energy as no transverse confinement was included.

\begin{figure}[htbp]
\begin{center}
   \includegraphics[width=\columnwidth]{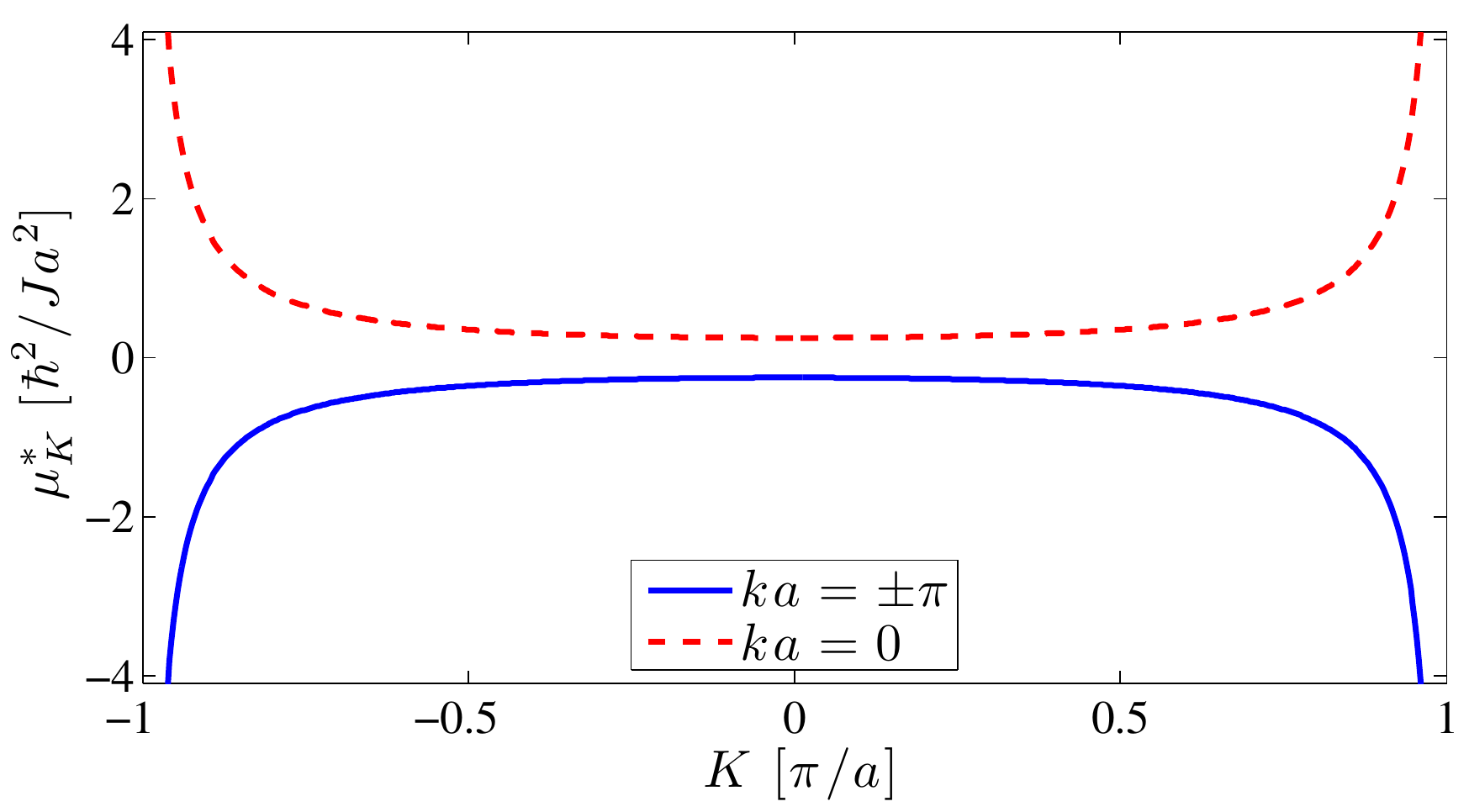} 
\caption{(Color online) The effective reduced mass at the upper band edge (solid line) and the lower band edge (dashed line) as a function of the center of mass momentum.}
\label{fig:m_eff}
\end{center}
\end{figure}

As $E_b$ approaches either of the band edges, the binding energy vanishes in a particularly simple way. To see this we observe that $E_b^2-E_K^2\approx  2\, {\rm{sgn}}(E_b)|E_K|E_{\rm{bind}}$ for $E_{\rm{bind}}\rightarrow 0$. Replacing $E_b$ with ${\rm{sgn}}(E_b)|E_K|$ in the denominator on the right hand side in (\ref{E_bind}) it then follows that
\begin{equation}
  \label{eq:E_bind_approx}
  E_{\rm{bind}} \approx \frac{{\rm{sgn}}(E_b)U^2/2|E_K|}{\left[1+\frac{{\mathcal{W}}^2/U}{E_{\rm{res}}(B,K)-{\mathcal{W}}^2/U-{\rm{sgn}}(E_b)|E_K|} \right]^2},
\end{equation}
as the bound state energy approaches ${\rm{sgn}}(E_b)|E_K|$. The denominator is exactly the resonance shape of the generalized scattering length (\ref{a_1D}), and the binding energy therefore approaches the universal form
\begin{equation}
E_{\rm{bind}} = - \frac{\hbar^2}{2\mu^*_K(a^\pm_{1{\rm{D}}})^2},
\label{E_bind_universal}
\end{equation}
when the scattering length is tuned to be large with respect to the lattice spacing $a$ in the vicinity of the Feshbach resonance. We have introduced the effective {\textit{reduced}} mass
\begin{equation}
\mu^*_K \equiv \hbar^2 \left( \frac{\partial^2 \epsilon_K}{\partial k^2} \right)^{-1}, 
\label{m_eff}
\end{equation}
of the atom pair with center of mass quasi-momentum, $K$, in the
lattice. Its limits at the top and the bottom of the band are $\mp
\hbar^2/|E_K|a^2$, with the upper (lower) sign applying for
$k\rightarrow \pm \pi/a$ ($k\rightarrow 0$). These limits are plotted in Fig.~\ref{fig:m_eff} as a function of $K$. From the expression for the effective reduced mass it is clear that the sign of the
universal binding energy expression (\ref{E_bind_universal}) adheres to the definition above.
The criteria of applicability for the quadratic approximation is
$|E_{\rm{bind}}| \ll {\rm{min}}(|E_K|,{\mathcal{W}}^2/U)$, or
$|\delta^{\pm}| \ll \sqrt{2|E_K|}{\mathcal{W}}^{3}/|U|^{5/2}$. The universal expression (\ref{E_bind_universal}) is compared with the exact binding energies in Fig.~\ref{fig:a_1D}(b) both above and below the continuum. Due to the proximity of the zero crossing of $a_{\rm{1D}}^-$, where (\ref{E_bind_universal}) diverges, the agreement below the band is worse than above the band.  

From Eqs. (\ref{a_1D}) and (\ref{E_bind_universal}) it is clear that the binding energy vanishes when $E_{\rm{res}}={\mathcal{W}}^2/U\pm|E_K|$. This condition for the disappearance of the Feshbach bound state coincides with the critical resonance energy at which the Fano profile develops a point of total transmission, as anticipated above.

Experimentally the binding energy of the dimer states can be measured by modulating the depth of the lattice. The bound pairs dissociate, when the frequency of modulation satisfies a resonance condition, such that the dimers can exchange energy with the undulating lattice. The resonance is marked by the appearance of unbound atoms~\cite{Winkler2006}.

\subsection{Bound state wavefunctions}
\label{sec:bound-state-wavefunction}

The formal solution of the bound state problem consistent with the single-pole approximation for the closed channel Green's function is 
\begin{equation}
\left(
\begin{array}{c}
|\psi^{\rm{op}}_b\rangle \\
|\psi^{\rm{cl}}_b\rangle \\
\end{array}
\right)
=
\sqrt{Z_b} \left(
\begin{array}{c}
\hat{G}^U_K(E_b,z)\hat{W}|\phi_{\rm{res}}\rangle \\
|\phi_{\rm{res}}\rangle
\end{array}
\right),
\end{equation}
where the closed channel population, $Z_b$, is determined by the normalization condition: $\langle \psi^{\rm{op}}_b|\psi^{\rm{op}}_b\rangle + \langle \psi^{\rm{cl}}_b|\psi^{\rm{cl}}_b\rangle=1$.
In coordinate space the open channel component of the bound state wavefunction therefore becomes
\begin{equation}
\psi_b^{\rm{op}}(z)=\sqrt{Z_b}{\mathcal{W}}G^U_K(E_b,z), 
\label{eq:psi_b_op}
\end{equation}
while the closed channel component of the bound state is described by
$\psi_b^{\rm{cl}}(z)=\sqrt{Z_b}\delta_{z,0}$. Here we have used that
the coupling is of zero range, and that the bare resonance state is
localized on a single lattice site.  The closed channel weight of a
bound atom pair with energy $E_b$ is
\begin{equation}
Z_b = \left[ 1+\frac{{\mathcal{W}}^2}{[1-UG^0_K(E_b,0)]^2}\frac{|E_b|}{(E_b^2-E_K^2)^{3/2}}\right]^{-1},
\label{Z}
\end{equation}
which is plotted in Figs~\ref{fig:R_vs_K}(i)-(l) and~\ref{fig:R_vs_Eres}(b).
As the bound state approaches the upper or the lower band edge, it follows from the limiting form of the binding energy that $E_b^2-E_K^2\approx E_K^2(a/a^{\pm}_{\rm{1D}})^2$, and consequently $Z_b$ vanishes as
$(U/{\mathcal{W}})^2|a/a^{\pm}_{\rm{1D}}|$ in the limit where $|a/a^{\pm}_{1{\rm{D}}}|\rightarrow 0$.  

In momentum space the open channel part of the bound state wavefunction is given by 
\begin{equation}
\phi_b^{\rm{op}}(k) = \sqrt{\frac{aZ_b}{2\pi}}{\mathcal{W}}\frac{{\mathcal{G}}_K^0(E_b,k)}{1-UG_K^0(E_b,z=0)},
\label{phi_mom}
\end{equation}
which is peaked near the center of the Brillouin zone ($k=0$) when the bound state energy is negative, and peaked near the edges of the Brillouin zone ($k=\pm \pi/a$) for $E_b>0$, making it possible to discern the bound states with $E_b<0$ from those with $E_b>0$ in time of flight imaging~\cite{Winkler2006}. This reflects that $\phi_K^{\rm{op}}(k)$ is predominantly comprised of the $k$-states with energies, $\epsilon_K(k)$, in proximity of $E_b$. For $E_b>0$ ($E_b<0$) this corresponds to the states near the top (bottom) of the band. 
The height of the peaks is reduced as $K$ approaches $\pm\pi/a$, where all relative motion quasi-momentum states are degenerate in our model. The closed channel component of the bound state is independent of the relative momentum: 
$\phi_b^{\rm{cl}}(k)=\sqrt{aZ_b/2\pi}$.

\begin{figure}[htbp]
\begin{center}
   \includegraphics[width=\columnwidth]{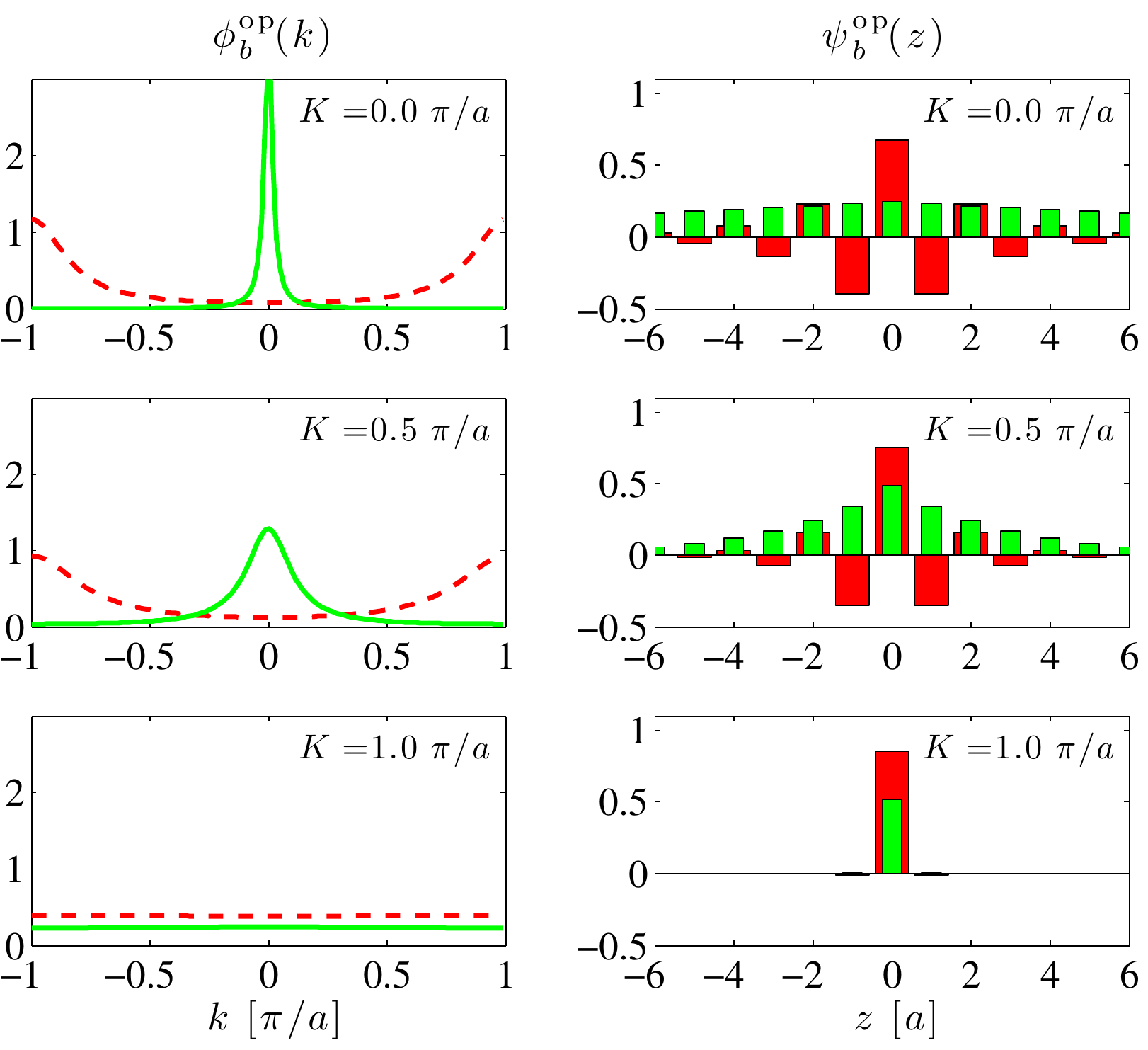} 
\caption{(Color online) Open channel components of bound state wavefunctions for $\bar{E}_{\rm{res}}=-J$, where a weakly bound molecular state (solid line, narrow bars) at rest is situated just below the band. The repulsively bound pair state lying above the band (dashed line, thick bars), is peaked at the edge of the Brillouin zone, and its sign alternates from one site to the next. The first column shows the momentum space wavefunction (\ref{phi_mom}) in units of $\sqrt{a/2\pi}$, the second column $\psi^{\rm{op}}_b(z)$.}
\label{fig:bound_state_wf}
\end{center}
\end{figure}

The size of the bound state 
\begin{equation}
\langle |z| \rangle=\frac{a}{2}\frac{Z_b{\mathcal{W}}^2}{[1-UG_K^0(E_b,0)]^2}\frac{E_K^2}{(E^2_b-E^2_K)^2}, 
\end{equation}
reflects the magnitude of the binding energy of atom pairs. In particular, it diverges when the bound state enters the continuum from above or from below: 
\begin{equation}
  \label{eq:norm_z_universal}
  \langle |z|\rangle \rightarrow \frac{|a^{\pm}_{\rm{1D}}|}{2} \ \ \ {\rm{for}} \ |a^{\pm}_{\rm{1D}}|\rightarrow \infty.
\end{equation}
In this limit the open channel part of the bound state wavefunction assumes a universal form
\begin{equation}
\psi^{\rm{op}}_b(z) \rightarrow \sqrt{\left|\frac{a}{a^{\pm}_{\rm{1D}}}\right|} e^{-|z/a^{\pm}_{\rm{1D}}|}\left[-{\rm{sgn}}(E_b)\right]^{z/a}.
\end{equation}

The bound state wavefunctions are shown in Fig.~\ref{fig:bound_state_wf} in both coordinate and momentum space for $\bar{E}_{\rm{res}}=-J$. At this resonance energy a bound molecular state exists just below the band for $K=0$ (lower bound state branch in Fig.~\ref{fig:R_vs_K}(b)). With a small binding energy, this state extends over several lattice sites as shown in Fig~\ref{fig:bound_state_wf}. For the repulsively bound pair state lying above the band (upper bound state branch in Fig.~\ref{fig:R_vs_K}(b)) the sign of the coordinate space wavefunction alternates between lattice sites.   

\subsection{Sweep experiments}

We have previously shown that a molecule has a finite probability of surviving a sweep through the continuum band when the applied magnetic field is varied across the Feshbach resonance~\cite{Nygaard2008}. The survival probability increases the faster the sweep and the weaker the coupling to the continuum. 
Given the existence of motionally bound states, an alternative experiment to investigate the passage of a molecule through the continuum may then be conducted by starting with the applied magnetic field held at a value such that $4J<|E_{\rm{res}}-{\mathcal{W}}^2/U|$, and molecules at rest prepared in the bound state with energy below $\bar{E}_{\rm{res}}$ (represented by the the lower branch of bound states in Figs.~\ref{fig:R_vs_K} and~\ref{fig:R_vs_Eres}). Accelerating the lattice imparts momentum $K_i$ to the molecules. If the magnetic field is then ramped adiabatically to a new value, where $|E_{K_i}|<|E_{\rm{res}}-{\mathcal{W}}^2/U|<4J$ the moving molecules remain stable, while molecules at rest are now subject to dissociative decay. With a further acceleration of the lattice, either in the opposite direction, or in the same direction, but letting $K$ cross one or several Brillouin zones, the molecules may be swept through the continuum to a final state with center of mass momentum $K_f$, where the molecules are stable provided $|E_{K_f}|<|E_{\rm{res}}-{\mathcal{W}}^2/U|$. The probability of a molecule surviving the center of mass momentum ramp through the continuum depends on the rate of lattice acceleration and the strength of the coupling to the continuum states.

\section{Spectral analysis}
\label{sec:spectral-analysis}

In this section we will perform a spectral analysis based on the full
Green's functions of both the open channel atoms and closed channel molecules. In general, given a retarded Green's
operator, $\hat{G}(E)$, the corresponding spectral function is
the imaginary part of the diagonal
Green's function, $A(E,b) = -2\text{Im} \langle b|\hat{G}(E)|b \rangle
$, in a basis $|b\rangle$ of orthonormal states. 
%The general form of the spectral function 
%\begin{equation}
%  \label{eq:spect_gen}
%  A(E,b)=2\pi|\langle  b|\Psi_{\eta}\rangle|^2D(E)  
%\end{equation}
%follows from the spectral decomposition of $\hat{G}(E)$.  Here $D(E)$
%and $|\Psi_{\eta}\rangle$ are the density of states (density of states) and the wave
%function of the interacting system, respectively, and $\eta$ is the set of quantum numbers for which $E(\eta)=E$. When $E$ is outside
%the band, $D(E)$ is a $\delta$-function in the energies of the bound
%states. The spectral function therefore directly provides the
%probability distribution of the bound states.

\subsection{The unperturbed case}
\label{sec:unperturbed-case}

For the unperturbed system ($U$=${\mathcal{W}}$=0) the spectral function is
\begin{equation}
  \label{eq:5}
  A^0_K(E,z)=2\pi D^0_K(E) \Theta(|E_K|-|E|), 
\end{equation}
where the density of states is
$D^0_K(E)=|\mathrm{d}\epsilon_K(q)/\mathrm{d}q|^{-1}=
-G^0_K(E,0)/i\pi$ for $|E|<|E_K|$.
The unperturbed system has no bound states, and hence the spectral
function is zero at all energies outside the continuum band. Inside the
band the scattering states are equally distributed over the whole
lattice, making the spectral function translationally invariant and
directly proportional to the density of states, which diverges at
the band edges, $E=\pm|E_K|$.

\subsection{The single-channel case}
\label{sec:non-resonant-case}

Both for the single-channel and the two-channel case we will need the entrance channel
Green's function, $G_K^U(E,z,z')\equiv \langle z|\hat{G}^U_K(E)|z'\rangle$, \textit{i.e.}, the propagator from one relative
coordinate in the lattice, $z'$, to another, $z$. When we consider just a single scattering channel this can be obtained from the Dyson equation
(\ref{Dyson}):
\begin{eqnarray}
  G^U_K(E;z,z') &=& G^0_K(E, z-z')\nonumber\\
    & & + \frac{G_K^0(E,z)U G_K^0(E,-z')}{1-UG_K^0(E,0)},
\end{eqnarray}
where we have used that the unperturbed Green's function is translationally invariant. 
This expression for $G_K^U(E,z,z')$ reduces to Eq. (\ref{GU_z}) when $z$ or $z'$ is equal to zero. 

For the spectral function we have to treat $|E|<|E_K|$ and $|E|>|E_K|$
separately.  Outside the continuum $G^0_K(E,z)$ is real and we obtain
the spectral function
\begin{equation}
  A^U_K(E,z) 
  %&=& 2\pi\delta(1-UG^0_K(E,0))
 % UG^0_K(E,0)^2e^{-2\kappa|z|}\nonumber\\
  = 2\pi \frac{Ue^{-2\kappa^0_b|z|}}{E^0_b} \delta(E-E^0_b),
\label{eq:A_U_bound}
\end{equation}
where $\kappa^0_b a=\cosh^{-1}|E^0_b/E_K|$ is the decay constant for the bound state. This gives the probability distribution
for finding a bound state atom pair $z$ lattice sites apart, when the bound state
energy is $E^0_b$. 

Inside the band we find
\begin{equation}
  \label{eq:9}
  A^U_K(E,z)=\pi\left(|\psi^{\rm{bg}}_K(E,z)|^2+|\psi^{\rm{bg}}_K(E,-z)|^2\right)D^0_K(E).
\end{equation}
where $\psi_K^{\rm{bg}}(E,z)$ is the scattering wave defined in
Eq. (\ref{psi_E_bg}). Changing the sign of $z$ is equivalent to
changing the direction of the incoming wave in the scattering wave
from $k$ to $-k$. These two waves are degenerate and therefore it is
natural to expect the average of the two for a given energy in the
spectral function.  We have used that the density of states is unchanged for elastic
scattering, \textit{i.e.} $D^0_K(E)$ is also the density of states for atoms scattering under
the influence of $\hat{U}$.

It is instructive to calculate the total spectral weight residing
inside the continuum band:
\begin{equation}
\int^{|E_K|}_{-|E_K|}  \frac{dE}{2\pi}\ A^U_K(E,z) = 1-\frac{Ue^{-2\kappa^0_b|z|}}{E^0_b}.
\end{equation}
With the addition of the bound state contribution (\ref{eq:A_U_bound})
the sum rule, $\int dE\ A^U_K(E,z) /2\pi = 1$, is then
seen to be satisfied.

\subsection{The two-channel case}
\label{sec:resonant-case}

When we have two coupled channels there is a spectral density for both
the open channel atoms and the closed channel molecules. 
We start by addressing the latter.

\subsubsection{The closed channel}
\label{sec:closed-channel}

\begin{figure}[htbp]
\begin{center}
   \includegraphics[width=\columnwidth]{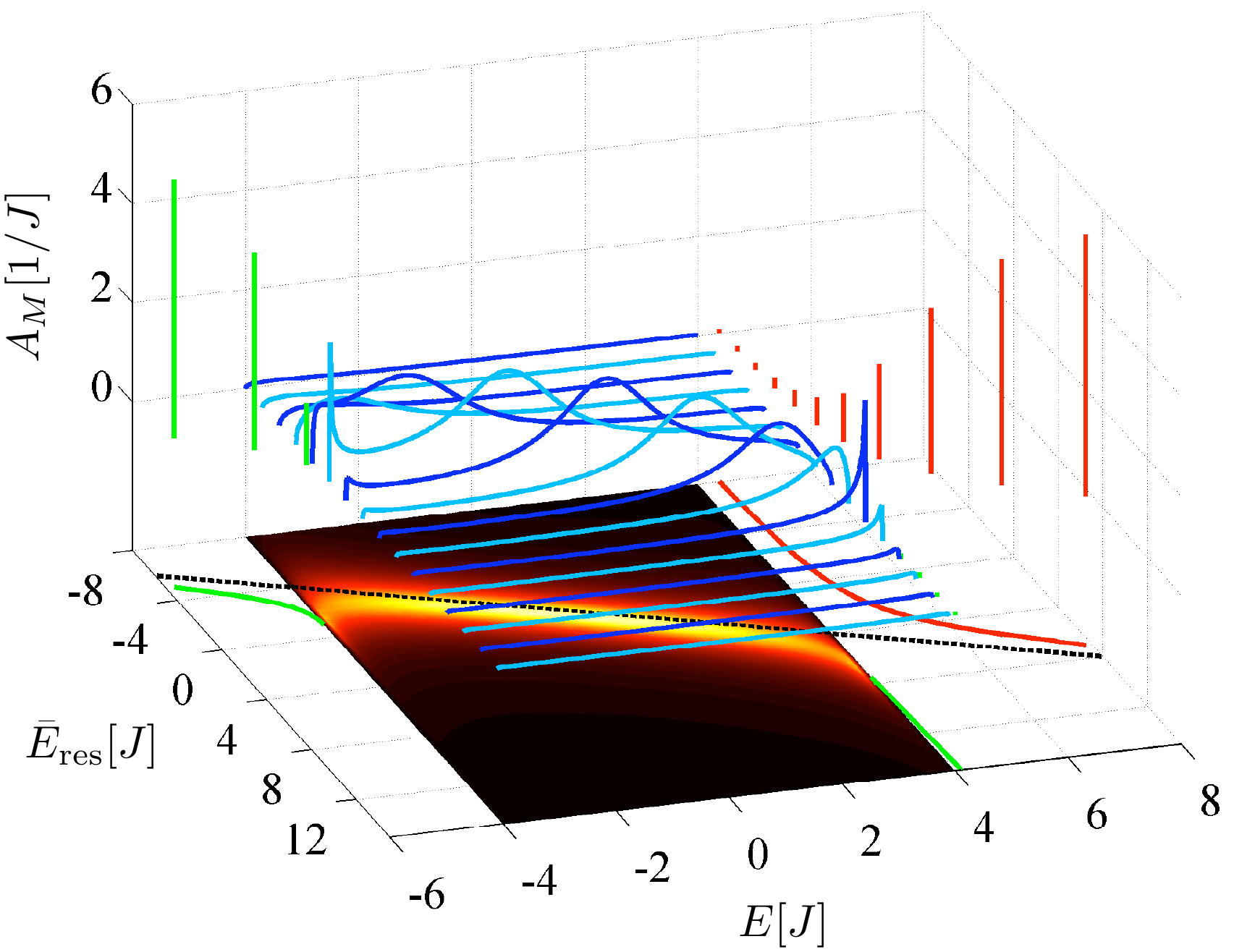} 
\caption{(Color online) Molecular spectral function as a function of the resonance
  energy for $K=0$. Inside the band $A_M$ is a continuous function given by (\ref{eq:A_M_continuum}). Outside the band the delta function peaks at the bound state poles of the propagator, $G_M$, are indicated by vertical lines with a height of $2\pi Z_b$. Offset below is a contour plot of $A_M$ inside the band, with bright colors indicating high spectral density. The solid lines in this plane are the bound state energies, while $\bar{E}_{\rm{res}}$ is indicated by the dashed line.}
\label{fig:A_M}
\end{center}
\end{figure}

The dressed Green's function for the closed channel is found by summing all the terms in the Dyson series
based on the bare propagator (\ref{G_cl_approx}). The result is
\begin{equation}
  G_M(E,K)= \frac{1}{E-E_{\rm{res}}(B,K)-\Sigma_M(E,K)},
\label{eq:G_M}
\end{equation}
where $\Sigma_M(E,K)=\mathcal{W}^2G^U_K(E,0)$ is the molecular
self-energy (\ref{sigma_cl}).  If $E$ is outside the continuum,
$\Sigma_M(E,K)$ is real, and the spectral function
is only non-zero at the poles of $G_M$, which coincide with those of the
scattering amplitudes, {\textit{i.e.}} the bound states of the
coupled channels problem (\ref{E_bind}).  The spectral weight at the
poles is given by the corresponding residues, which are just the probability of finding the
bound state pair in the closed channel, $Z_b$:
\begin{equation}
   A_M(E,K) = 2\pi Z_b\delta(E-E_b).
\label{eq:A_M_bound}
\end{equation}
For energies inside
the continuum, we obtain the usual structure,
\begin{equation}
 A_M(E,K) = \frac{\hbar\Gamma(E,K)}{\left[E-E_{\rm res}(B,K)-\Delta(E,K)\right]^2 + \frac{\hbar^2\Gamma^2(E,K)}{4}},
\label{eq:A_M_continuum}
\end{equation}
where the energy dependent shift, $\Delta$, and decay rate, $\Gamma$, are
defined in Eqs. (\ref{eq:delta}) and (\ref{eq:gamma}),
respectively. In the middle of the band the shift and width are slowly
varying functions of energy, and the molecular spectral function has
an approximately Lorentzian shape with a peak position given approximately by the resonance energy $E_{\rm{res}}$. However, near the band edges
significant deviations from this simple form arise
due to the strong interference with the continuum threshold. This leads to a double peak structure of $A_M$, when the bound state is about to enter the continuum from below, as is evident from Fig.~\ref{fig:A_M}, which shows $A_M$ as a function of the energy and $E_{\rm{res}}$ for a molecule at rest.  

The molecule lifetime also follows from Fermi's golden rule,
$\Gamma(E,K)=2\pi|\langle\phi_{\rm res}|\hat W|\psi^{\rm
  bg}_K\rangle|^2D^0(E)/\hbar$. Though the density of states diverges at the edges
of the continuum (van Hove singularities) the decay rate goes to zero
at both thresholds, due to vanishing overlap of the background
scattering state with the bare resonance state when $E=\pm |E_K|$.

\subsubsection{The open channel}

For the open channel the full Green's function in the resonant case reads
\begin{eqnarray}
  \label{eq:3}
    G_K(E;z,z') &=& G_K^0(E,z-z')\nonumber\\
    & &+\frac{G_K^0(E,z)\Sigma_K(E,B)G_K^0(E,-z')}{1-\Sigma_K(E,B)G_K^0(E,0)}
\end{eqnarray}
where $\Sigma_K(E,B)=U+{\mathcal{W}}^2/[E-E_{\rm{res}}(B,K)]$ is the
self-energy for the relative motion of an atom pair in the open channel. 
This gives the spectral function
\begin{equation}
  A^{\rm{op}}_K(E,z)=2\pi Z_b  [{\mathcal{W}}G^U_K(E,z)]^2\delta(E-E_b)
\end{equation}
for energies outside the band. 
The open channel Green's functions has exactly the same poles as the molecular propagator (\ref{eq:G_M}), but the residues at the atomic poles depend on the separation between the atoms in the pair. 
However, using the identity 
$\sum_{z\in {\mathbb{Z}}}e^{-2\kappa |z|}=EG_K^0(E,0)$
one may show that 
\begin{equation}
\sum_{z\in {\mathbb{Z}}}   A^{\rm{op}}_K(E,z)= 2\pi\left(1-Z_b\right) \delta(E-E_b),
\end{equation}
such that the combined spectral weight of the open channel and closed channel at a pole is unity. We must sum over all possible separations for the open channel weight, since the component of the bound state wavefunction in the open channel  is distributed over the lattice according to (\ref{eq:psi_b_op}).

\section{Inclusion of higher Bloch bands}
\label{sec:validity}

Thus far we have only investigated the dynamics in the lowest Bloch
band of the lattice potential. In this section we analyze the
neglected coupling to higher bands. First we estimate the size of the
second order energy shift from the excluded higher bands and derive a
criterion for the validity of our theoretical model.  Next, in
Sec. \ref{sec:incl-high-bloch} we discuss how a modest coupling to
higher atomic bands is expected to modify the central conclusions of
this work.

\subsection{Validity of the lowest band approximation}
\label{sec:criterion-usage}

In this section we set up a criterion for using the theory, by
comparing the second order energy shift from the omitted atomic and
molecular exited Bloch bands to the relevant energy scales.
The dominant contribution to the energy shift comes from the bare
resonance state coupling to atom pairs in higher bands, while the
shift from open channel atoms coupling to higher molecular Bloch bands
are several orders of magnitude smaller for two reasons.
Firstly, the coupling matrix elements between an atom pair in the lowest band and 
an excited band bare molecule are much smaller than the coupling to a molecule in the lowest band 
(see appendix~\ref{sec:mol-couplings}). Secondly, the energy splitting between the ground and excited molecular bands is larger than the atomic band gap, since the molecules experience a deeper lattice potential. Hence we approximate the
unperturbed wavefunction by $|\phi_{\rm res}\rangle$, the resonance state in the lowest molecular band. For simplicity we will
furthermore assume the width of the Bloch bands to be small compared
to the separation from the lowest band, so that the energy can be
approximated by the average energy of the band.

Under these assumptions the second order energy shift is
\begin{equation}
  \Delta E^{(2)} =
  -\sum_{(n,m)\neq (1,1)}\frac{|\mathcal{W}_{1nm}|^2}{E_{n}+E_m-2E_{1}},
\label{eq:2nd_order_shift}
\end{equation}
where the sum is over pairs of atoms in the $n$'th and $m$'th Bloch
band, $\mathcal{W}_{1nm}$ is the inter-band coupling matrix element defined in
appendix \ref{sec:calc-param}, and $E_{n}$ is the average energy of the
$n$'th band. If the atoms are identical bosons, we can take the sum to
be over $n\leq m$ to avoid double counting.

We will require the shift to be small compared to the relevant energy
scales. If $\Delta E^{(2)}$ is comparable to the gap between the ground
and first excited band $2(E_{2}-E_{1})$, the higher bands will become
populated, and the lowest band approximation of our model will break
down~\cite{Kohl2005}. On the other hand a stronger condition can be derived by
requiring that the energy shift be small in comparison with
the width of the ground band, $8J$, which sets the relevant energy scale for the dynamics in our model. By fitting the respective energies from numerical band structure calculations at different lattice depths we arrive at the condition
\begin{equation}
  \label{eq:validity_condition}
  \bigg(\frac{a_{\rm bg}}{100 \ a_{\rm B}}\bigg)\bigg(\frac{\Delta\mu}{\mu_{\rm{B}}}
  \bigg)\bigg(\frac{\Delta B}{1{\rm{G}}}
  \bigg)\bigg(\frac{m}{\rm amu}
  \bigg)\bigg(\frac{\lambda_{L}}{1000\rm nm}\bigg)\ll
  \frac{Q(V_0^{\parallel})}{\big(V_0^\perp\big)^{0.6}},
\end{equation}
where $Q(V_0^\parallel)=30(V_0^\parallel+4)$ when comparing $\Delta E^{(2)}$ with the band gap, and $Q(V_0^\parallel)=60\exp(-0.2V_0^\parallel)$ when requiring the energy shift to be much smaller than the width of the lowest band~\cite{Note1}. 
The scaling factors are the Bohr
radius, $a_{\rm{B}}$, the Bohr magneton, $\mu_{\rm B}$, Gauss (G), the atomic mass unit ($\rm amu$), and nanometers. If both
criteria are met, coupling to higher bands can be safely neglected.
Otherwise, if the energy shift incurred from the coupling to atoms in higher
Bloch bands is significant with respect to the band width yet negligible with regards to the band gap, we might expect quantitative changes to our results, but as we discuss below, the qualitative features of our model will remain. In the case where the lowest band approximation fails coupling to higher bands can be systematically accounted for in the tight binding limit~\cite{Dickerscheid2005,Gubbels2006,Diener2006}, leading to an effective single-band Hamiltonian with a non-local  coupling between dressed molecules and unbound atom pairs~\cite{Duan2005}.

Both of the above stated conditions can be fulfilled for the Feshbach
resonance studied by Syassen \emph{et al.} in $^{87}$Rb near 414
G~\cite{Syassen2007}. In the entrance channel both atoms are in the $|F=1,m_F=0\rangle$ hyperfine state. 
The resonance is narrow, with $\Delta B=18$~mG,
and has a very small difference in magnetic moments between the two
channels, $\Delta \mu = 111$~kHz/G, which is preferable if
Eq. (\ref{eq:validity_condition}) is to hold.  

A final condition for our results to be experimentally relevant is that the variation of the resonance energy
due to magnetic field noise $\delta B_{\rm{rms}}$ is much smaller than the
bandwidth $8J$.  
With a lattice wavelength of 830.44~nm~\cite{Syassen2007} the recoil energy is $E_{\rm{R}}=3.5$~kHz
(equivalent to $0.16 \ \mu{\rm{K}}$) and for the particular resonance considered here
$E_{\rm{R}}/\Delta\mu=$32 mG.  For
a longitudinal lattice depth of $E_{\rm{R}}$ we therefore require $\delta
B_{\rm{rms}}\ll 46 \ {\rm{mG}}$ according to Table~\ref{tab:parameters}. In Ref. \cite{Durr2004} the magnetic
field noise was found to be less than 4 mG.

\subsection{Corrections due to higher atomic Bloch bands}
\label{sec:incl-high-bloch}

As discussed in the previous subsection the major correction to our
model comes from the coupling to higher atomic bands.  We now include
these couplings into our model. For each atomic band we apply the
analysis of Sec. \ref{sec:entr-chann-greens} to obtain band specific
Green's functions $G^0_{K,n,m}(E,z)$ and $G^U_{K,n,m}(E,z)$ for the relative motion of one
atom in Bloch band $n$ and  another in band $m$. As for the first
band $\hat{G}^0_K(E)$ is the Green's function for noninteracting particles and
$\hat{G}^U_K(E)$ is the Green's function for particles with an on-site
interaction, $\hat{U}$. We determine $G^0_{K,n,m}(E,z)$ and $G^U_{K,n,m}(E,z)$ by
replacing $\epsilon_K(k)$ by
$[-2J_{n1}\cos(Ka/2+ka)-2J_{m1}\cos(Ka/2-ka)+E_n+E_m-2E_1]$ in
Eq. \eqref{G0_k},
and replacing $U$ by the band specific interaction strength $U_{nm}$
in Eq. \eqref{GU_z} (for the definition of $U_{nm}$ see Appendix
\ref{sec:calc-param}).

We now introduce the coupling $\hat{W}$ between the first molecular
band and all the atomic bands. This gives the molecular self-energy
\begin{eqnarray}
  \Sigma_M(E,K)&=&\sum_{n,m}\langle \phi_{\rm{res}}|\hat{W}\hat{G}^U_{K,n,m}(E)\hat{W}|\phi_{\rm{res}}\rangle\nonumber\\
  &=&\sum_{n,m}|\mathcal{W}_{1nm}|^2G^U_{K,n,m}(E,0),
\label{eq:Sigma_M_higher}
\end{eqnarray}
which reduces to (\ref{sigma_cl}), when we exclude contributions from
higher bands.  Because we only consider energies well outside the
excited bands, $G^U_{K,n,m}$ is real and slowly varying with
energy for $n,m\neq 1$. Therefore we conclude that the lifetime
$\Gamma$ of the molecules is unchanged and that the change in the
resonance shift $\Delta={\rm Re} \Sigma_M$ must vary slowly. A change
of the resonance shift by an amount $\Delta'$ will affect some of the
results in Secs. \ref{sec:two-chann-scatt}, \ref{sec:bound-states} and
\ref{sec:resonant-case}. The changes amount to simply replacing
$E_{\rm res}$ by $E_{\rm res}+\Delta'$, a renormalized resonance
energy.

For $E=0$ and $K=0$ we obtain for the parameters in \cite{Syassen2007} an approximate relationship for the
change in resonance shift
\begin{equation}
  \label{eq:DeltaChange}
  \Delta' \simeq -0.004 E_{\rm R} (V_0^\parallel)^{-0.1} (V_0^\perp)^{0.6}
\end{equation}
due to the higher atomic bands~\cite{Note2}.  For the lattice
depth used throughout this paper $\Delta'$ is only 2\% of the
bandwidth $8J$, and it is reasonable to neglect higher bands.

A direct coupling between ground and excited atomic bands may be included by renormalizing the local interactions to include the effect of virtual transitions to excited levels through band changing collisions. This leads to an energy dependent correction to the coupling parameters~\cite{Gurvitz1993}. But since the correction is inversely proportional to the gap separating the ground and excited bands, it is also expected to be small.  

We have assumed onsite interactions and only nearest-neighbor tunneling
in all atomic bands which is a questionable assumption for the higher
bands, but Eq. (\ref{eq:DeltaChange}) still provides an estimate of
the magnitude of the correction to the real part of the molecular self-energy.

\section{Concluding remarks}
\label{sec:Discussion}

We have presented a complete description of the Feshbach physics of an atom pair in an optical lattice represented by a discrete two-channel Hamiltonian. We have demonstrated how the structured continuum due to the periodic potential modifies the resonance line shape and position as well as the properties of the bound states outside the continuum. 

The lattice potential facilitates repulsively bound pair states in the gap between the two-particle Bloch bands, and near a Feshbach resonance a further class of exotic bound states may be created, which are only stable when their center of mass momentum exceeds a critical value, thus enabling us to tune the Feshbach physics by controlling the molecular motion. Ignoring the small modulation of the resonance energy due to tunneling of the bare molecules, the Feshbach molecules are stable bound states of the system if their center of mass quasi-momentum in the first Brillouin zone satisfies $|K|>K_c$, where 
\begin{equation}
K_c=\frac{2}{a}\cos^{-1}\left(\frac{|\bar{E}_{\rm{res}}(B)-{\mathcal{W}}^2/U|}{4J}\right).
\end{equation}   
If $|K|<K_c$ the pair state lies inside the continuum and decays with a rate $\Gamma(E,K)$, which we have determined analytically.  For magnetic fields such that $|\bar{E}_{\rm{res}}(B)-{\mathcal{W}}^2/U|>4J$ the Feshbach molecules are stable for all $K$. 
Note that while the decay rate near threshold is proportional to the density of states, which diverges at the band edges, $\Gamma(E,K)$ goes continuously to zero as $E$ approaches the edge of the continuum, since the overlap between the bare resonance state and the open channel continuum states is suppressed due to the background interactions in the open channel. 

An additional effect of the background interactions is to shift and modify the resonance profile. With a non-zero onsite interaction $U$ the position of the resonance is shifted away from the bare resonance energy $E_{\rm{res}}(B,K)$, and the transmission of atoms in the open channel is described by a Fano profile, with a shape determined by the energy dependent shift and width of the resonance. The energy dependence is most pronounced near the band edges, where the coupling to the continuum is strongest. This gives rise to threshold effects, when the resonance energy is tuned close to the band edge, as evidenced e.g. in the double peak structure of the molecule spectral function. We have outlined how the energy dependence of the scattering can be mapped out spectroscopically by inducing dissociative transitions from molecular bound states to two-particle scattering states inside the band. 

From the expansion of the scattering amplitude as the collision energy approaches the band edges we have defined a generalized scattering length, which exhibits a Feshbach resonance at both the lower and the upper threshold of the continuum. The diverging scattering length coincides with the appearance/disappearance of the Feshbach bound state.  In the limit of a deep lattice potential the lowest Bloch band narrows, and these two resonances coalesce. Since the bare molecules and the entrance channel atoms experience different lattice induced energy offsets, the position of both the upper and the lower threshold Feshbach resonance is shifted relative to the position of the Feshbach resonance in free space. For the specific resonance we have considered here, this shift exceeds the width of the resonance. 

\appendix

\section{Interaction parameters}
\label{sec:calc-param}

For the interaction between the atoms in the open channel we assume that $\hat{U}$ is a
contact potential
%, $\langle {\bf x}|\hat{U}|{\bf x}'\rangle=g_{\rm{bg}}\delta({\bf x}-{\bf x}')\delta({\bf x})$, 
of strength $g_{\rm bg}=4\pi\hbar^2 a^{3\rm{D}}_{\rm bg}/m$, where $a^{3\rm{D}}_{\rm bg}$ is the (free space) background scattering length.  
Therefore the on-site interaction between a pair of atoms in
bands $n$ and $m$ becomes
\begin{eqnarray}
  \label{eq:U_mn}
    U_{nm}&=&g_{\rm bg}
\prod_{i=1,2}
    \int dx_i \,
      \big[w_{10}^\perp(x_i)\big]^4
      \nonumber \\
    &&\times\int dx_3 \, \big[ w_{n0}^\parallel(x_3)w_{m0}^\parallel(x_3)
    \big]^2,
\end{eqnarray}
under the assumption that only the lowest band in the transverse lattice is occupied. Since all lattice wells are identical,  
the integrals can be performed over Wannier functions centered on any site.

The coupling to the closed channel is of short range, since it is proportional to the difference between the atomic triplet and singlet potentials, which have the same long range form~\cite{Kohler2006}. In addition, the closed channel bound state has a size (typically tens of {\AA}), which is small compared with the lattice spacing (hundreds of nm).
We thus assume a purely on-site coupling, which in free space would couple molecular and atomic plane wave states with a  strength $g=\sqrt{4\pi\hbar^2 a^{3\rm{D}}_{\rm
    bg}\Delta\mu\Delta B/m}$. Here $\Delta\mu$ is the
difference between the closed and open channel magnetic moments and
$\Delta B$ is the magnetic field width of the Feshbach resonance in the absence of the lattice.  In the discrete lattice model this results
in the matrix element
\begin{eqnarray}
  \label{eq:W}
    \mathcal{W}_{lnm} &=& g \prod_{i=1,2}\int dx_i \, w^{\rm
        cl,\perp}_{10}(x_i)w_{10}^\perp(x_i)^2   \nonumber \\
    &&\times\int dx_3 \, w^{\rm
      cl,\parallel}_{l0}(x_3)w_{n0}^\parallel(x_3)w_{m0}^\parallel(x_3)
\end{eqnarray}
for the coupling of a pair of atoms in bands $n$ and $m$ to a bare
molecule in band $l$. The matrix elements are zero whenever $l+n+m$ is
even due to parity of the Wannier functions. A molecule experiences
twice as deep a lattice as the atoms, see Eq. (\ref{V_lat_mol}), and
has twice the atomic mass. Consequently, separate Wannier functions,
$w^{\rm{cl}}_{nj}(x_i)$, have to be calculated for the center of mass
motion of the resonance state.

With the exception of 
Sec.~\ref{sec:validity} we will consider exclusively the lowest Bloch band for the atoms and molecules.
Therefore, we will only need three parameters to characterize the
dynamics in the lattice: $J=J_{11}$, $U=U_{11}$ and
$\mathcal{W}=\mathcal{W}_{111}$.

As discussed in section~\ref{sec:validity} our discrete lattice model is valid for a narrow Feshbach resonance. As a particular example we use the 
resonance in $^{87}$Rb near 414
G, which has been studied by Syassen \emph{et
  al.}~\cite{Syassen2007}. For this resonance $\Delta B=18$ mG and
$\Delta \mu = 111$ kHz/G, while the background scattering length is
100.8 Bohr radii. For all the plots in this paper we consider a
lattice with $V_0^\perp=30$ and $V_0^\parallel=1$. This results in
interaction parameters $U/J=1.4$ and
${\mathcal{W}}/J=2.2$. Table~\ref{tab:parameters} gives the relevant
model parameters for this resonance and a range of lattice depths.  

\begin{table}
\begin{tabular}{cccccc}
\hline
\hline
$V_0^\parallel$ & $V_0^\perp$ & $J/E_{\rm{R}}$ & $J_m/E_{\rm{R}}$ & $U/J$ & ${\mathcal{W}}/J$ \\ 
\hline
1 & 15 & 0.18  & 0.043 & 0.9 & 1.8 \\
2 & 15 & 0.14  & 0.015 & 1.4 & 2.5 \\ 
3 & 15 & 0.11  & 0.006 & 2.1 & 3.4 \\
4 & 15 & 0.085 & 0.003 & 3.0 & 4.6 \\
5 & 15 & 0.066 & 0.001 & 4.2 & 6.2 \\
1 & 30 & 0.18  & 0.043 & 1.4 & 2.2 \\
2 & 30 & 0.14  & 0.015 & 2.1 & 3.0 \\
3 & 30 & 0.11  & 0.006 & 3.1 & 4.1 \\
4 & 30 & 0.085 & 0.003 & 4.5 & 5.6 \\
5 & 30 & 0.066 & 0.001 & 6.3 & 7.6 \\
\hline
\hline
\end{tabular}\caption{Discrete lattice parameters for the $^{87}$Rb Feshbach resonance near 414 G. The recoil energy is $E_{\rm{R}}=3.5$ kHz.}
\label{tab:parameters}
\end{table}

\section{Free space resonance shift}
\label{Resonance_shift}

In free space the coupling between the collision channels induces a shift from the magnetic field $B^{\rm{free}}_{\rm{res}}$, where the resonance state is degenerate with the entrance channel threshold  [$E^{\rm{free}}_{\rm{res}}(B^{\rm{free}}_{\rm{res}})=0$], to the position $B^{\rm{free}}_0$ of the scattering length divergence. This offset may be approximately evaluated by
\begin{equation}
B^{\rm{free}}_0-B^{\rm{free}}_{\rm{res}} = \Delta B f(y).
\end{equation}
Here $f(y)=y(1-y)/[1+(1-y)^2]$ is a dimensionless function of the ratio $y=a^{3\rm D}_{\rm{bg}}/\bar{a}$ between the background (free space) scattering length and the so-called mean scattering length $\bar{a}=(mC_6/\hbar^2)^{1/4}\Gamma(3/4)/2^{3/2}\Gamma(5/4)\approx 0.478(mC_6/\hbar^2)^{1/4}$, which is a characteristic length scale of the long range $-C_6/r^6$ part of the molecular potential~\cite{Kohler2006}. 
In the lattice the energy of the closed channel resonance state relative to the zero of energy (center of the lowest band) is then given by
\begin{equation}
E_{\rm{res}}(B)=\Delta\mu[B-B_0^{\rm{free}}+\Delta Bf(y)]+E_m(K)-E_a,
\end{equation}
where the lattice induced energy shifts $E_a$ and $E_m(K)$ of the open and closed channel, respectively, are defined in Fig.~\ref{fig:EnergyDiagram}.

\section{Coupling to higher molecular bands}
\label{sec:mol-couplings}

The molecular Wannier
states, $\prod_{i=1,2,3}w^{{\rm{cl}},i}_{n_ij_i}(x_i)$, where $n_i$
and $j_i$ refers to the molecular Bloch band index and lattice site
index along direction $i$, respectively, constitute a complete set on
which we can expand the atomic function $f({\bf x})
=\prod_{i=1,2}[w_{10}^\perp(x_i)^2]\times
w_{n0}^\parallel(x_3)w_{m0}^\parallel(x_3)$. If we denote the expansion
coefficients $c^{j_1j_2j_3}_{n_1n_2n_3}$, we have
$\mathcal{W}_{lnm}=gc^{000}_{11l}$ and $U_{nm}=g_{\rm bg}||f||^2$, where $||\cdot||^2$ is the $L^2$-norm. Now
a simple norm argument gives the following identity for the neglected
molecular couplings to higher bands
\begin{equation}
  g^2\!\!\!\!\!\!\!\!\!\!\!\!\!\! \sum_{\left(\substack{j_1j_2j_3\\n_1n_2n_3}\right)
    \neq\left(\substack{000\\111}\right)}
  \!\!\!\!\!\!\!\!\!\!\!\!\!\!
  |c^{j_1j_2j_3}_{n_1n_2n_3}|^2=\frac{g^2U_{nm}}{g_{\rm bg}}-|\mathcal{W}_{1nm}|^2,
\label{eq:neglected_coupling}
\end{equation}
where $g^2/g_{\rm bg}=\Delta\mu\Delta B=0.6E_R$ for the 414 G
resonance in $^{87}$Rb. For a lattice with $V_0^{\parallel}=1$ and $V_0^{\perp}=30$ this results in the estimate $\sum_{l\neq
  1}|\mathcal{W}_{l11}|^2\leq 0.0059 \mathcal{W}_{111}^2$ for an atom pair in the lowest band ($n=m=1$). We therefore neglect the coupling of ground band atoms to molecules in higher bands in comparison with the matrix elements, which couple molecules in the lowest Bloch band with excited atom pairs.

\acknowledgments{We thank Blair Blakie for suggesting the useful
  connection between the binding energy and the effective mass. N. N. acknowledges financial support by the Danish Natural Science Research Council.}


\begin{thebibliography}{90}

\bibitem{Dahan1996} M. Ben Dahan, E. Peik, J. Reichel, Y. Castin, and C. Salomon, Phys. Rev. Lett. {\bf{76}} 4508 (1996).
\bibitem{Wilkinson1996} S. R. Wilkinson, C. F. Bharucha, K. W. Madison, Q. Niu, and M. G. Raizen, Phys. Rev. Lett. {\bf 76}, 4512 (1996).
\bibitem{Greiner2002} M. Greiner, M. O. Mandel, T. Esslinger, T. H{\"a}nsch, and I. Bloch, Nature (London) {\bf 415}, 39 (2002).
\bibitem{Bloch2007} I. Bloch, J. Dalibard, and W. Zwerger, Rev. Mod. Phys. {\bf{80}}, 885 (2008).
\bibitem{Mandel2003} O. Mandel, M. Greiner, A. Widera, T. Rom, 
T. W. H{\"a}nsch, and I. Bloch , Nature (London) {\bf 425}, 937 (2003).
\bibitem{Kohler2006} T. K{\"o}hler, K. G{\'o}ral, and P. S. Julienne, Rev. Mod. Phys. {\bf{78}}, 1311 (2006).
\bibitem{Thalhammer2006} G. Thalhammer, K. Winkler, F. Lang, S. Schmid, R. Grimm, and J. Hecker Denschlag, Phys. Rev. Lett. {\bf 96}, 050402 (2006).
\bibitem{Volz2006} T. Volz, N. Syassen, D. M. Bauer, E. Hansis, S. D{\"u}rr, and G. Rempe, Nature Phys. {\bf 2}, 692 (2006).
\bibitem{Winkler2006}  K. Winkler, G. Thalhammer, F. Lang, R. Grimm, J. Hecker Denschlag, A. J. Daley, A. Kantian, H. P. B{\"u}chler, and P. Zoller, Nature (London) {\bf{441}}, 853 (2006).
\bibitem{Ospelkaus2006} C. Ospelkaus, S. Ospelkaus, L. Humbert, P. Ernst, K. Sengstock, and K. Bongs, Phys. Rev. Lett. {\bf 97}, 120402 (2006).
\bibitem{Veiga2002} P. A. F. da Veiga, L. Ioriatti, and M. O'Carroll, Phys. Rev. E
{\bf 66}, 016130 (2002). 
\bibitem{Mahajan2006} S. M. Mahajan and A. Thyagaraja, J. Phys. A {\bf 39}, L667 (2006).
\bibitem{Denschlag2006} J. Hecker Denschlag and A. J. Daley,
  Proceedings of the international school of physics ``Enrico Fermi'',
  Course CLXIV, Ultra-Cold Fermi Gases, e-print arXiv:cond-mat/ 
0610393. 
\bibitem{Grupp2007} M. Grupp, R. Walser, W. P. Schleich, A. Muramatsu, and M. Weitz, J. Phys. B. {\bf{40}}, 2703 (2007).
\bibitem{Nygaard2006} N. Nygaard, B. I. Schneider, and P. S. Julienne, Phys. Rev. A {\bf{73}}, 042705 (2006).
\bibitem{Fano1961} U. Fano, Phys. Rev. {\bf{124}}, 1866 (1961).
\bibitem{Anderson1961} P. W. Anderson, Phys. Rev. {\bf{124}}, 41 (1961).
\bibitem{Duan2005} L.-M. Duan, Phys. Rev. Lett. {\bf{95}}, 243202 (2005).
\bibitem{Carr2005} L. D. Carr and M. J. Holland, Phys. Rev. A {\bf{72}}, 031604(R) (2005).
\bibitem{Zhou2005} F. Zhou, Phys. Rev. B {\bf{72}}, 220501(R) (2005).
\bibitem{Nygaard2008} N. Nygaard, R. Piil, and K. M{\o}lmer, Phys. Rev. A {\bf 77}, 021601(R) (2008).
\bibitem{Foelling2007} S. F{\"o}lling, S. Trotzky, P. Cheinet, M. Feld, R. Saers, A. Widera, T. M{\"u}ller, and I. Bloch, Nature (London) {\bf 448}, 1029 (2007).
\bibitem{Strohmaier2007} N. Strohmaier, Y. Takasu, K. G{\"u}nter, R. J{\"o}rdens, M. K{\"o}hl, H. Moritz, and T. Esslinger, Phys. Rev. Lett {\bf 99}, 220601 (2007).
\bibitem{Syassen2007} N. Syassen, D. M. Bauer, M. Lettner, D. Dietze, T. Volz, S. D{\"u}rr, and G. Rempe, Phys. Rev. Lett. {\bf{99}}, 033201 (2007). 
\bibitem{Nockel1994} J. U. N{\"o}ckel and A. D. Stone, Phys. Rev. B {\bf{50}}, 17415 (1994).
\bibitem{Olshanii1998} M. Olshanii, Phys. Rev. Lett. {\bf{81}}, 938 (1998).
\bibitem{Bergeman2003} T. Bergeman, M. G. Moore, and M. Olshanii, Phys. Rev. Lett. {\bf{91}}, 163201 (2003).
\bibitem{Peano2005} V. Peano, M. Thorwart, C. Mora, and R. Egger, New. J. Phys. {\bf{7}}, 192 (2005).
\bibitem{Kim2005} J. I. Kim, J. Schmiedmayer, and P. Schmelcher, Phys. Rev. A {\bf{72}}, 042711 (2005).
\bibitem{Yurovsky2005} V. A. Yurovsky, Phys. Rev. A {\bf 71}, 012709 (2005).
\bibitem{Yurovsky2006a} V. A. Yurovsky, Phys. Rev. A {\bf 73}, 052709 (2006). 
\bibitem{Piil2007} R. Piil and K. M{\o}lmer, Phys. Rev. A {\bf{76}}, 023607 (2007).
\bibitem{Gurvitz1993} S. A. Gurvitz and Y. B. Levinson, Phys. Rev. B {\bf 47}, 10578 (1993).
\bibitem{Eberly1965} J. Eberly, Am. J. Phys. {\bf{33}}, 771 (1965).
\bibitem{Moerdijk95} A. J. Moerdijk, B. J. Verhaar, and A. Axelsson,
  Phys. Rev. A {\bf 51}, 4852 (1995).
\bibitem{Orso2005} G. Orso, L. P. Pitaevskii, S. Stringari, and M. Wouters, Phys. Rev. Lett. {\bf 95}, 060402 (2005).
\bibitem{Micnas1990} R. Micnas, J. Ranninger, and S. Robaszkiewicz, Rev. Mod. Phys. {\bf 62}, 113 (1990). 
\bibitem{Kuklov2003} A. B. Kuklov and B. V. Svistunov, Phys. Rev. Lett. {\bf 90}, 100401 (2003).
\bibitem{Duan2003} L.-M. Duan, E. Demler, and M. D. Lukin, Phys. Rev. Lett {\bf 91}, 090402 (2003).
\bibitem{Petrosyan2007} D. Petrosyan, B. Schmidt, J. R. Anglin, and M. Fleischhauer, Phys. Rev. A {\bf 76}, 033606 (2007).
\bibitem{Miyakawa2007} T. Miyakawa and P. Meystre, J. Low. Temp. Phys {\bf 148}, 429 (2007).
\bibitem{Kohl2005} M. K{\"o}hl, H. Moritz, T. St{\"o}ferle, K. G{\"u}nter, and T. Esslinger, Phys. Rev. Lett. {\bf{94}}, 080403 (2005).
\bibitem{Note1} The fit was performed for $1\leq V_0^\parallel \leq 15$ and $10\leq V_0^\perp \leq 35$, and the sum in (\ref{eq:2nd_order_shift}) was taken over the lowest 35 bands.  
\bibitem{Dickerscheid2005} D. B. M. Dickerscheid, U. Al Khawaja, D. van Oosten, and H. T. T. C. Stoof, Phys. Rev. A {\bf{71}}, 043604 (2005).
\bibitem{Gubbels2006} K. B. Gubbels, D. B. M. Dickerscheid, and H.  T. C. Stoof, New. J. Phys {\bf{8}}, 151 (2006).
\bibitem{Diener2006} R. B. Diener and T.-L. Ho, Phys. Rev. Lett. {\bf{96}}, 010402 (2006).  
\bibitem{Durr2004} S. D{\"u}rr, T. Volz, and G. Rempe, Phys. Rev. A {\bf{70}}, 031601(R) (2004).
\bibitem{Note2} This fit was performed for $1\leq V_0^\parallel \leq 35$ and $14\leq V_0^\perp \leq 35$, and the contribution from the first 35 bands were included in (\ref{eq:Sigma_M_higher}).  

\end{thebibliography}
\end{document}